\documentclass[12pt]{article}
\usepackage{amsmath}
\usepackage{amssymb}
\usepackage{latexsym}
\usepackage{rotate}
\usepackage[dvips]{epsfig}
\title{Derivation and Empirical Validation of
a Refined Traffic Flow Model}
\author{Dirk Helbing\\[3mm]
{\protect\em II. Institute of Theoretical Physics}\\
{\protect\em University of Stuttgart}\\
{\protect\em 70550 Stuttgart}\\
{\protect\em Germany}}
\begin{document}
\maketitle
\vfill
\begin{abstract}
The gas-kinetic foundation of fluid-dynamic traffic equations suggested
in previous papers [Physica A {\bf 219}, 375 and 391] is further refined
by applying the theory of dense gases and granular materials to the
Boltzmann-like traffic model by Paveri-Fontana.
It is shown that, despite the phenomenologically similar behavior of
ordinary and granular fluids, the relations for these cannot
directly be transferred to vehicular traffic. The dissipative and anisotropic
interactions of vehicles as well as their velocity-dependent space
requirements lead to a considerably different structure of the
macroscopic traffic equations, also in comparison with the previously
suggested traffic flow models. As a consequence, the instability
mechanisms of emergent density waves are different. 
Crucial assumptions are validated by empirical traffic
data and essential results are illustrated by figures.\\[4mm]
\end{abstract}
%\pacs{47.50.+d,51.10.+y,47.55.-t,89.40.+k} % 83.10.Hh
\clearpage
\section{Introduction} \label{intro}

\begin{figure}[htbp]
\unitlength1cm
\begin{center}
\begin{picture}(14.2,10.6)(-0.2,-0.8)
\put(0,9.8){\epsfig{height=14cm, angle=-90, 
      bbllx=50pt, bblly=50pt, bburx=554pt, bbury=770pt, % clip=,
      file=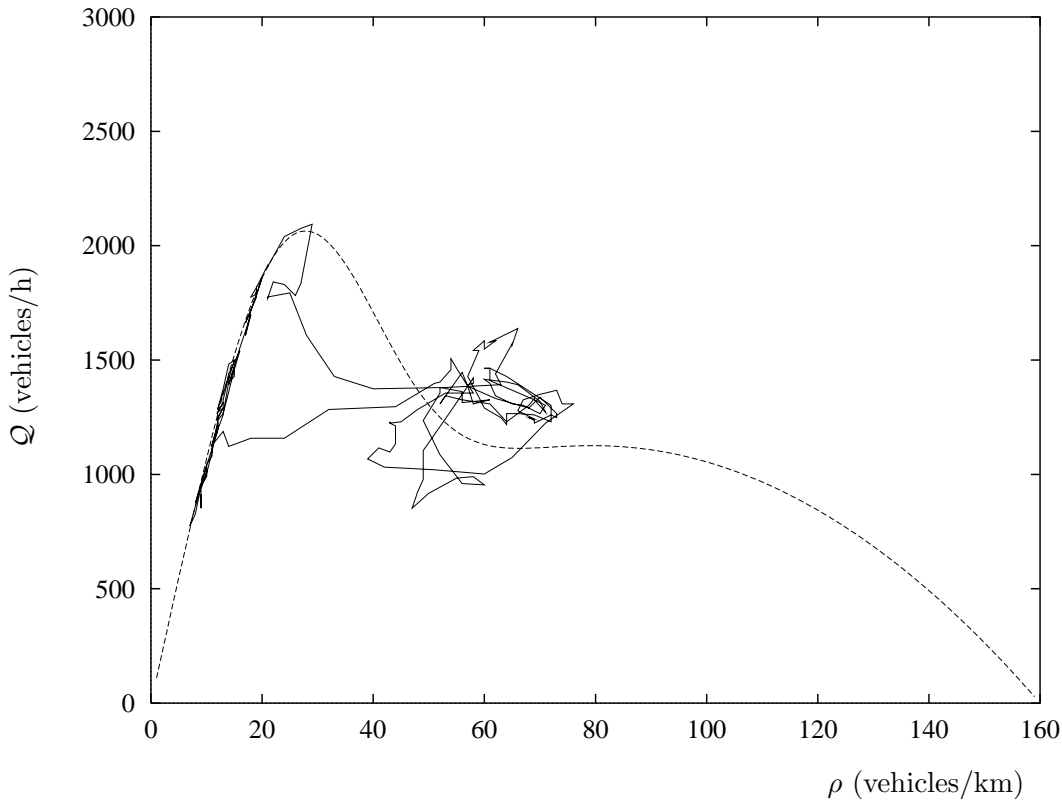}}
\put(12,-0.6){\makebox(0,0){\small $\rho$ (vehicles/km)}}
\put(0,5.1){\makebox(0,0){\rotate[l]{\hbox{\small ${\cal Q}$ (vehicles/h)}}}}
\end{picture}
\end{center}
\caption[]{Illustration of the {\protect\em fundamental diagram} ${\cal Q}(\rho)
= \rho V_{\rm e}(\rho)$ describing the equilibrium flow-density relationship 
(--~--) and of the temporal course ${\cal Q}(r,t) = \rho(r,t)V(r,t)$ 
of the empirical flow (---). Obviously, the equilibrium relation
${\cal Q}(r,t) = {\cal Q}_{\rm e}(\rho(r,t))$ is only fulfilled at small
densities. After the maximum flow is reached, the
temporal development of the flow ${\cal Q}(r,t)$ shows the typical
hysteresis phenomenon discovered by Hall \protect\cite{Hall} which indicates
a phase transition from almost homogeneous flow to stop-and-go traffic.
(Empirical data: Cross-section 
$r = 41.8$\,km of the Dutch two-lane highway A9 from Haarlem to Amsterdam
at November 2, 1994, between 6:30\,am and 10:00\,am.)}
\label{hysteresis}
\end{figure}

\begin{figure}[htbp]
\unitlength1cm
\begin{center}
\begin{picture}(14,10.6)(0,-0.8)
\put(0,9.8){\epsfig{height=14cm, angle=-90, 
      bbllx=50pt, bblly=50pt, bburx=554pt, bbury=770pt, % clip=,
      file=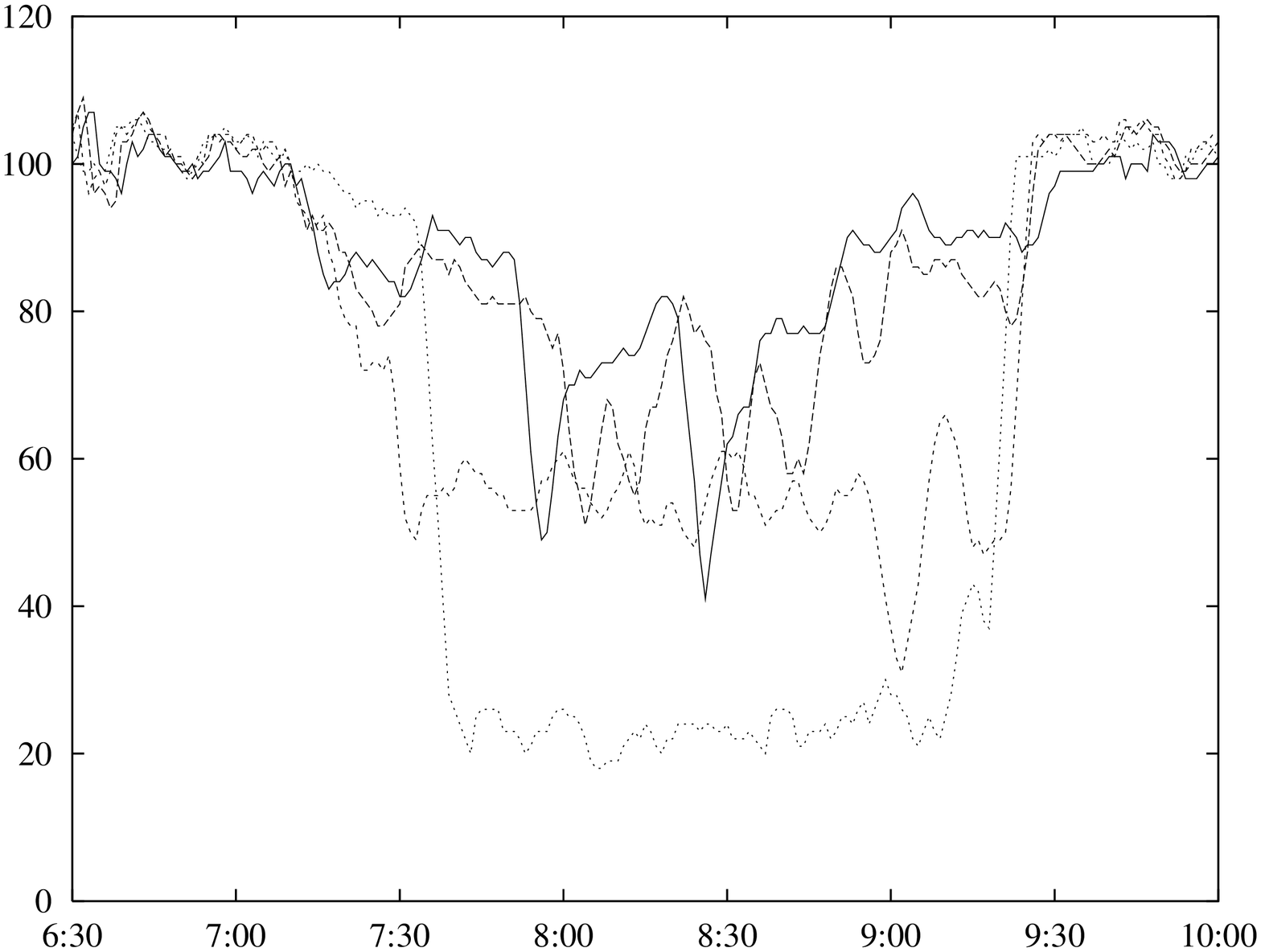}}
\put(13,-0.6){\makebox(0,0){\small $t$ (h)}}
\put(0.2,5.1){\makebox(0,0){\rotate[l]{\hbox{\small $V(r,t)$ (km/h)}}}}
\end{picture}
\end{center}
\caption[]{Temporal evolution of the mean velocity $V(r,t)$ at subsequent
cross-sections of the Dutch highway A9 from Haarlem to Amsterdam
at October 14, 1994 (five minute averages of single vehicle data). 
The prescribed speed limit is 120\,km/h. We observe
a breakdown of velocity during the rush hours between 7:30\,am and
9:30\,am due to the overloading of the highway at
$r = r_0 := 41.8$\,km ($\cdots$). %Because of the off-ramp at $r_0$,
At the subsequent cross-sections the traffic situation recovers 
(-~-~-: $r = r_0+1$\,km; --~--: $r = r_0+2.2$\,km; ---: $r = r_0+4.2$\,km).
Nevertheless, the amplitudes of the small velocity fluctuations at $r_0$
become larger and larger, leading to so-called stop-and-go waves, i.e.\ to
alternating periods of decelerating and accelerating traffic.} 
%At the same time, the wave lengths of the stop-and-go waves become larger.} 
\label{stopandgo}
\end{figure}

Modelling of traffic flow on highways presently attracts a rapidly growing
community of physicists. Recent research focusses on microsimulation
models \cite{mic} as well as on fluid-dynamic Navier-Stokes-like
models \cite{Hel,Hill,K,KK} and their derivation from the ``microscopic''
behavior of driver-vehicle units via gas-kinetic equations
\cite{Gas,Nels}. Of particular interest is the description of traffic 
instabilities (cf.\ Fig.\ \ref{hysteresis})
which lead to the emergence of stop-and-go traffic above
a certain critical vehicle density. This phenomenon is illustrated by
Figure \ref{stopandgo}. 
%\begin{center}
%{\em Insert Figures \ref{hysteresis} and \ref{stopandgo} about here.}
%\end{center}
\par
\begin{figure}[htbp]
\unitlength1cm
\begin{center}
\begin{picture}(14,10.6)(0,-0.8)
\put(0,9.8){\epsfig{height=14cm, angle=-90, 
      bbllx=50pt, bblly=50pt, bburx=554pt, bbury=770pt, % clip=,
      file=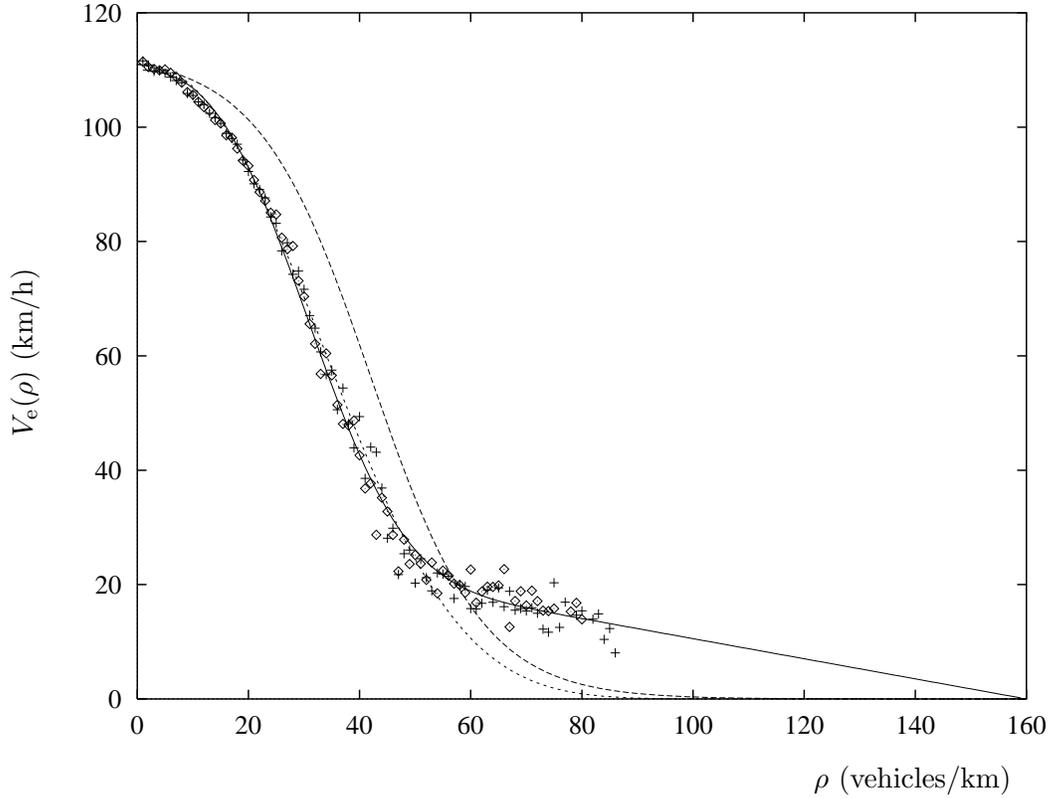}}
\put(12,-0.6){\makebox(0,0){\small $\rho$ (vehicles/km)}}
\put(0.2,5.1){\makebox(0,0){\rotate[l]{\hbox{\small $V_{\rm e}(\rho)$ (km/h)}}}}
\end{picture}
\end{center}
\caption[]{Illustration of different suggestions for the
equilibrium velocity-density relation $V_{\rm e}(\rho)$
(---: Helbing \protect\cite{Hab};
--~--: Kerner and Konh\"auser \protect\cite{KK};
-~-~-: K\"uhne \protect\cite{Kvel} and Cremer \protect\cite{Cvel}),
compared with empirical traffic data 
of the Dutch highway A9 from Haarlem to Amsterdam, where a speed limit
of $V_0 = 120$\,km/h applies ($\Diamond$: October 14, 1994;
+: November 2, 1994). The empirical data are mean values of all
one minute averages which fall into the range $[\rho-0.5$\,veh/km,
$\rho+0.5$\,veh/km). $V_{\rm e}(0) < V_0$ is because of the law that 
speedometers must show the actual speed or more, but never less.}
\label{eqvel}
\end{figure}

In section \ref{models} it will be shown that most proposed macroscopic 
traffic models can be viewed as special cases of
the {\em continuity equation}
\begin{equation}
 \frac{\partial \rho}{\partial t} + \frac{\partial {\cal Q}}{\partial r} = 0
 \qquad \mbox{with} \qquad {\cal Q} = \rho V
\label{conteq}
\end{equation}
for the spatial {\em density} $\rho(r,t)$ per lane and
and a {\em velocity equation} of the form
\begin{equation}
 \frac{\partial V}{\partial t} + V \frac{\partial V}{\partial r}
 = - \frac{1}{\rho} \frac{\partial {\cal P}}{\partial r}
 + \frac{1}{\tau} [V_{\rm e}(\rho) - V] \, .
\label{veleq}
\end{equation}
Here, $V\partial V/\partial r$ is the so-called {\em convection term
(transport term)} which describes velocity changes arising from 
the motion with mean velocity $V$. The term containing the {\em 
traffic pressure} ${\cal P}$ is the {\em anticipation term} and takes into
account that driver-vehicle units react to the traffic situation 
in front of them. The {\em relaxation term} $(V_{\rm e}-V)/\tau$ reflects
the adaptation of the {\em mean velocity} $V(r,t)$ to the 
density-dependent {\em equilibrium
velocity} $V_{\rm e}(\rho)$ (cf.\ Fig.~\ref{eqvel})
with a {\em relaxation time} $\tau$.  
\par
Apart from the relaxation term, the continuity equation and the 
velocity equation have the form of the fluid-dynamic equations for
compressible gases. The additional relaxation term reflects that
the mean velocity of vehicles decreases at bottlenecks, in contrast to
ordinary fluids which flow faster. Later on, it will turn out that the
relaxation term is responsible for the emergence of stop-and-go traffic.
\par
It should be remarked that all macroscopic traffic models of the
above type base on phenomenological reasoning in a number of points. 
Therefore, they imply at least one of the following inconsistencies
\cite{Dag,Nels,Hel,Gas}, depending on the respective model (details later):
\begin{itemize}
\item The variation of individual vehicle velocities is neglected.
Some models take it into account by the dependence ${\cal P} = \rho\Theta$ of
the traffic pressure ${\cal P}$ on the velocity variance $\Theta$.
\item The %equilibrium relation for the 
velocity variance does not decrease with growing density.
\item The %equilibrium 
variance does not vanish when the mean velocity
vanishes. Therefore, large pressure gradients,
which can cause the development of negative velocities, are possible
(cf.\ Eq.\ (\ref{veleq})).
%, if the mean velocity is very small.
\item Vehicles are implicitly treated as point-like objects.
\item For a certain density range,
the traffic pressure decreases with growing density.
This is connected with an acceleration of vehicles into regions with 
larger density (cf.\ Eq.\ (\ref{veleq})), 
so that drivers would race into existing traffic jams. 
\item In the course of time, the density $\rho(r,t)$ can, on certain
conditions, exceed the bumper-to-bumper density $\rho_{\rm bb} = 1/l_0$,
where $l_0$ denotes the average vehicle length.
\item The region of unstable traffic flow, which is related with stop-and-go
traffic, is not correctly described. In reality,
traffic is only stable at very 
small densities (free flow) and extreme densities (slow-moving traffic).
\item Emerging density waves develop a shock-like (i.e. almost discontinuous)
structure.
\item The viscosity terms which are implicitly (by the integration method)
or explicitly introduced for a numerical solution
of the macroscopic traffic equations have no theoretical foundation. 
\end{itemize}
In this paper we will show,
how to derive a consistent model from a suitable gas-kinetic
(Boltzmann-like) model. %(cf.\ Sections \ref{imp} and \ref{hightraf}). 
For this purpose, we will extend Paveri-Fontana's Boltzmann-like model 
(cf.\ Sec.\ \ref{PaFo}) by diffusion effects
due to imperfect driving (cf.\ Sec.\ \ref{imp}). Moreover, we have to
take into account the space requirement by vehicles. This will be
done analogously to the theory of dense gases and granular flows \cite{densgas}
(cf.\ Sec.\ \ref{high}). 
It will turn out that the structure of the resulting fluid-dynamic
traffic equations changes considerably compared to Eqs.\ (\ref{conteq}) and
(\ref{veleq}), as well as compared to the equations for dense gases or
granular materials (cf.\ Sec.\ \ref{highgran}). Reasons for this are the loss of
momentum conservation, the anisotropy of vehicular interactions, and
the velocity-dependence of vehicular space requirements. Consequently,
a direct transfer of relations for ordinary gases or fluids as proposed 
by previous traffic models is not possible.

\section{Historical evolution of macroscopic traffic models}\label{models}

From equations (\ref{conteq}) and (\ref{veleq}) we obtain the traffic models
suggested by other researchers, if ${\cal P}$, $V_{\rm e}$, and $\tau$
are specified in a suitable way:
\begin{itemize}
\item In the limit $\tau \rightarrow 0$ we have
\begin{equation}
 V(r,t) \approx V_e(\rho(r,t)) \, .
\end{equation}
The resulting model
\begin{equation}
 \frac{\partial \rho}{\partial t}
 + \left( V_{\rm e} + \rho \frac{\partial V_{\rm e}}{\partial \rho}
 \right) \frac{\partial \rho}{\partial r} = 0
\label{LWR}
\end{equation}
was independently proposed by Lighthill and Whitham \cite{LW} 
and Richards \cite{Rich}. It describes the formation of {\em kinematic waves}
which propagate with velocity 
\begin{equation}
 c(\rho) = \rho \frac{\partial V_{\rm e}}{\partial \rho}
\end{equation}
relative to $V_{\rm e}$. Nevertheless, the model cannot descibe
the emergence of stop-and-go traffic, since wave amplitudes are not
amplified. The wave profile only becomes steeper and steeper in the course 
of time, so that it builds up a shock-like structure (i.e.\ it
becomes discontinuous). Consequently, a simulation of Eq.\ (\ref{LWR})
is very problematic. Numerically treatable variants of the above model
were developed by Daganzo \cite{Daganz}.
\item Payne \cite{Payne} suggested the model
\begin{equation}
  \frac{\partial V}{\partial t} + V \frac{\partial V}{\partial r} \equiv
 \frac{dV}{dt} = - \frac{1}{\tau} \left( V - V_e - 
 \frac{\partial \rho/\partial r}{2\rho}\frac{\partial V_e}{\partial
   \rho} \right)
 = - \frac{\nu}{\rho\tau} \frac{\partial \rho}{\partial r}
 + \frac{1}{\tau} ( V_{\rm e} - V )
\end{equation}  
with $\nu = \frac{1}{2} |\partial V_{\rm e}/\partial \rho|$,
corresponding to the traffic pressure 
\begin{equation}
 {\cal P} = - V_e/(2\tau) \, .
\end{equation}
He derived his model from a microscopic follow-the-leader model \cite{FL}
by means of a Taylor expansion \cite{Payne}. This model was also used
by Papageorgiou \cite{Papa}. Cremer \cite{Cremer} slightly modified the
anticipation term by a factor $\rho/(\rho + \varkappa)$. Finally, Smulders
\cite{Smulders} introduced additional fluctuation terms in the continuity and
velocity equation. However, the resulting equations still
predict the development of shock-like density changes, if no {\em ``numerical
viscosity''} is introduced.
%The problem of Payne's model is that it is either
%stable or unstable for all densities, depending on the value of the
%relaxation time $\tau$. As a consequence, it also does not correctly describe
%the emergence of stop-and-go waves above a certain critical density.
\item In the limit $\tau \rightarrow 0$, Payne's model implies
\begin{equation}
  V(r,t) = V_e(\rho(r,t)) - \frac{\tau}{\rho}\frac{\partial {\cal P}}
 {\partial \rho}\frac{\partial \rho(r,t)}{\partial r} \, .
\end{equation}
This results in an additional diffusion term to the
Lighthill-Whitham-Richards model,
\begin{equation}
  \frac{\partial \rho}{\partial t} + \left( V_e + \rho \frac{\partial
   V_e}{\partial \rho} \right)  \frac{\partial \rho}{\partial r}
 = {\cal D} \frac{\partial^2 \rho}
 {\partial r^2} \, ,
\label{plusdiff}
\end{equation}
which solves the shock-formation problem. The diffusion function is
\begin{equation}
 {\cal D} = \tau \frac{\partial {\cal P}}{\partial \rho} 
 = \frac{1}{2} \left| \frac{\partial V_{\rm e}}{\partial \rho} \right| \, .
\end{equation}
\item A discrete cell model which is very suitable for real-time 
simulations of large traffic networks was proposed by
Hilliges and Weidlich \cite{Hill}. This consists of the discrete
continuity equation
\begin{equation}
  \frac{\partial \rho(i,t)}{\partial t}
 + \frac{1}{\Delta r} [ {\cal Q}(i,t) - {\cal Q}(i-1,t) ] = 0 \, .
\label{basic}
\end{equation}
Due to the anticipatory driver behavior the flow is modelled by
\begin{equation}
 {\cal Q}(i,t) := \rho(i,t) V_e(\rho(i+1,t)) \, ,
\end{equation}
which assumes that drivers adapt to the velocity in the next cell.
Therefore, the cell length
$\Delta r \approx 100$\,m must be chosen in agreement with
the driver behavior. By Taylor approximation of the Hilliges-Weidlich
model, we again obtain equation (\ref{plusdiff}), but this
time the diffusion function is given by
\begin{equation}
  {\cal D} = \frac{\Delta r}{2} \left( V_e + \rho 
 \left|\frac{\partial V_e}{\partial \rho}\right| \right) \, .
\end{equation}
In the special case 
\begin{equation}
 V_{\rm e}(\rho) = V_0 \left( 1 - \frac{\rho}{\rho_{\rm max}} \right) \, ,
\end{equation}
equation (\ref{plusdiff}) can be transformed to the analytically solvable
Burgers equations \cite{Burgers}
\begin{equation}
 \frac{\partial C}{\partial t} + C(r,t) 
 \frac{\partial C}{\partial r} = {\cal D} \frac{\partial^2 C}{\partial
   r^2} 
\end{equation}
with 
\begin{equation}
  C(r,t) = V_0 \left( 1 - \frac{2\rho(r,t)}{\rho_{\rm max}} \right)    \, .
\end{equation}
Since the stationary solutions of Eq.\ (\ref{plusdiff}) are stable with respect
to fluctuations, the above model cannot describe the formation of 
stop-and-go waves. Therefore, an extension by a dynamic velocity equation of 
the form (\ref{veleq}) with ${\cal P} = 0$ was proposed \cite{Hill}.
\item An alternative approach to the model of Payne was
suggested by Phillips \cite{Phil}, who derived his model from a
Boltzmann-like traffic equation. For the traffic pressure he obtained
the gas-kinetic relation 
\begin{equation}
 {\cal P} = \rho \Theta \, ,
\end{equation}
where he assumed
\begin{equation}
 \Theta(\rho) = \Theta_0 \left( 1 - \frac{\rho}{\rho_{\rm max}} \right) \, ,
\end{equation}
since the {\em velocity variance} $\Theta$
should vanish at the {\em maximum traffic density}
$\rho_{\rm max}$. However, according to this formula the 
density-gradient $\partial {\cal P}/\partial \rho$ of the traffic pressure
will be negative in a certain density range (cf.\ Fig.~\ref{dpress}). 
\item K\"uhne \cite{K} as well as Kerner, and Konh\"auser \cite{KK} avoided
this problem by assuming $\Theta(\rho) = \Theta_0$. However, 
$\Theta$ cannot be interpreted as velocity variance, then, so that their
equations are not compatible with gas-kinetic traffic models. Moreover,
in order to smooth out developing shock structures, 
K\"uhne introduced an additional {\em viscosity term} $\tilde{\nu} \partial^2
V/\partial r^2$ \cite{K}. In analogy to the Navier-Stokes equations for 
ordinary fluids, Kerner and Konh\"auser assumed $\tilde{\nu}(\rho) 
= \eta_0/ \rho$, 
so that their model corresponds to the effective traffic pressure
\begin{equation}
  {\cal P} = {\cal P}_{\rm e} - \eta_0 \frac{\partial V}{\partial r} 
\qquad \mbox{with} \qquad {\cal P}_{\rm e} = \rho \Theta_0 \, .
\end{equation}
Simulations of their model show the formation of density clusters and
stop-and-go waves at moderate densities \cite{KK}. The instability
region can be determined by a {\em linear stability analysis} about the
stationary and spatially homogeneous solution $\rho(r,t) = \rho_{\rm e}$
and $V(r,t) = V_{\rm e}(\rho_{\rm e})$. Inserting the small overall
perturbation
\begin{eqnarray}
 \delta \rho(r,t) := \rho(r,t) - \rho_{\rm e} &=&
 \int dk \, \hat{\rho}(k) \exp[{\rm i}kr + (\lambda - {\rm i} \omega)t] \, ,
\nonumber \\
 \delta V(r,t) := V(r,t) - V_{\rm e}(\rho_{\rm e}) &=&
 \int dk \, \hat{V}(k) \exp[{\rm i}kr + (\lambda - {\rm i} \omega)t] 
\end{eqnarray}
into Eqs.\ (\ref{conteq}) and (\ref{veleq}), applying a Taylor expansion,
neglecting non-linear contributions,
and applying the orthogonality relations for the complex exponential functions
yields the following linear eigenvalue problem:
\begin{equation}
 \left( \begin{array}{ccc}
-\tilde{\lambda} &  &  - {\rm i} k \rho_{\rm e} \\ 
 & & \\
\displaystyle 
-  \frac{{\rm i} k}{\rho_{\rm e}} 
\frac{\partial {\cal P}_{\rm e}}{\partial \rho}
+ \frac{1}{\tau} \frac{\partial V_{\rm e}}{\partial \rho} & & \displaystyle
- \tilde{\lambda} - \frac{\eta_0 k^2}{\rho_{\rm e}} - \frac{1}{\tau}  
\end{array}\right) \left(
\begin{array}{c}
\hat{\rho}(k) \\
\mbox{ } \\
\hat{V}(k)
\end{array} \right)
\stackrel{!}{=} \left( \begin{array}{c}
0 \\
\mbox{ } \\
0
\end{array} \right) \, .
\label{eigen}
\end{equation}
Here, we have introduced the abbreviation
\begin{equation}
 \tilde{\lambda} := \lambda - {\rm i} \tilde{\omega} \qquad \mbox{with} \qquad
 \tilde{\omega} := \omega - k V_{\rm e}(\rho_{\rm e}) \, .
\end{equation}
$k$ is the {\em wave number}, $\lambda$ the {\em growth parameter},
and $\omega$ the {\em oscillation frequency} of the perturbations.
Eq.\ (\ref{eigen}) is fulfilled for the solutions of the {\em characteristic
polynomial}
\begin{equation}
 \tilde{\lambda}^2 + \tilde{\lambda} \left( \frac{\eta_0 k^2}{\rho_{\rm e}}
 + \frac{1}{\tau} \right) 
 + {\rm i}k\rho_{\rm e} \left( - \frac{{\rm i}k}{\rho_{\rm e}}
 \frac{\partial {\cal P}_{\rm e}}{\partial \rho} + \frac{1}{\tau}\frac{\partial
 V_{\rm e}}{\partial \rho} \right) = 0 \, .
\end{equation}
This leads to 
\begin{eqnarray}
 \tilde{\lambda} &=& - \frac{1}{2T} \pm \sqrt{ \frac{1}{4T^2}
 - ( C_{\rm r} + {\rm i} C_{\rm i} ) } 
 = - \frac{1}{2T} \pm \sqrt{\Re \pm {\rm i} |\Im | } \nonumber \\
 &=& - \frac{1}{2T} \pm
 \left[ \sqrt{\frac{1}{2}\left( \sqrt{\Re^2 + \Im^2} + \Re \right) }
 \pm {\rm i} \sqrt{\frac{1}{2}\left( \sqrt{\Re^2 + \Im^2} - \Re \right) }
 \right] \, ,
\end{eqnarray}
where we have defined
\begin{equation}
 \frac{1}{T} := \frac{\eta_0 k^2}{\rho_{\rm e}} + \frac{1}{\tau} \, , \qquad
 C_{\rm r} := k^2 \frac{\partial {\cal P}_{\rm e}}{\partial \rho} \, , \qquad
 C_{\rm i} := \frac{k \rho_{\rm e}}{\tau} \frac{\partial V_{\rm e}}
 {\partial \rho} \, ,
\end{equation}
and
\begin{equation}
 \Re := \frac{1}{4T^2} - C_{\rm r} = \frac{1}{4T^2} 
 - k^2 \frac{\partial {\cal P}_{\rm e}}{\partial \rho} \, , \qquad
 \pm |\Im| := - C_{\rm i} = \frac{k \rho_{\rm e}}{\tau} 
 \left| \frac{\partial V_{\rm e}}{\partial \rho} \right| \, . 
\end{equation}
A transition from stability to instability occurs on the condition
\begin{equation}
 \lambda = - \frac{1}{2T} \pm \sqrt{\frac{1}{2}
 \left( \sqrt{\Re^2 + \Im^2} + \Re \right) } \stackrel{!}{=} 0 
\end{equation}
which implies
\begin{equation}
 C_{\rm i} \stackrel{!}{=} \pm \frac{1}{T} \sqrt{C_{\rm r}} \, .
\end{equation}
Therefore, the equilibrium solution of the K\"uhne-Kerner-Konh\"auser
model is unstable on the condition
\begin{equation}
 \rho_{\rm e} \left| \frac{\partial V_{\rm e}}{\partial \rho} \right|
  > \sqrt{\frac{\partial {\cal P}_{\rm e}}{\partial \rho}} 
 \left( 1 + \frac{\tau \eta_0 k^2}{\rho_{\rm e}} \right) \, .
\label{instcond}
\end{equation}
This condition is fulfilled at moderate densities, where the equilibrium
velocity $V_{\rm e}$ rapidly decreases with growing density
(cf.\ Figs.\ \ref{eqvel} and \ref{kernkon}a). According to
the continuity equation (\ref{conteq}), this decrease of velocity 
causes a further increase of
density which finally leads to the formation of density clusters.
Short wave lengths (i.e.\ large wave numbers) are stable because of the
smoothing effect of viscosity $\eta_0$.
\par
\begin{figure}[htbp]
\unitlength1cm
\begin{center}
\begin{picture}(14,6.8)(0,-0.8)
\put(0,6){\epsfig{height=14cm, angle=-90, 
      bbllx=9cm, bblly=4.5cm, bburx=18.5cm, bbury=25cm, clip=,
      file=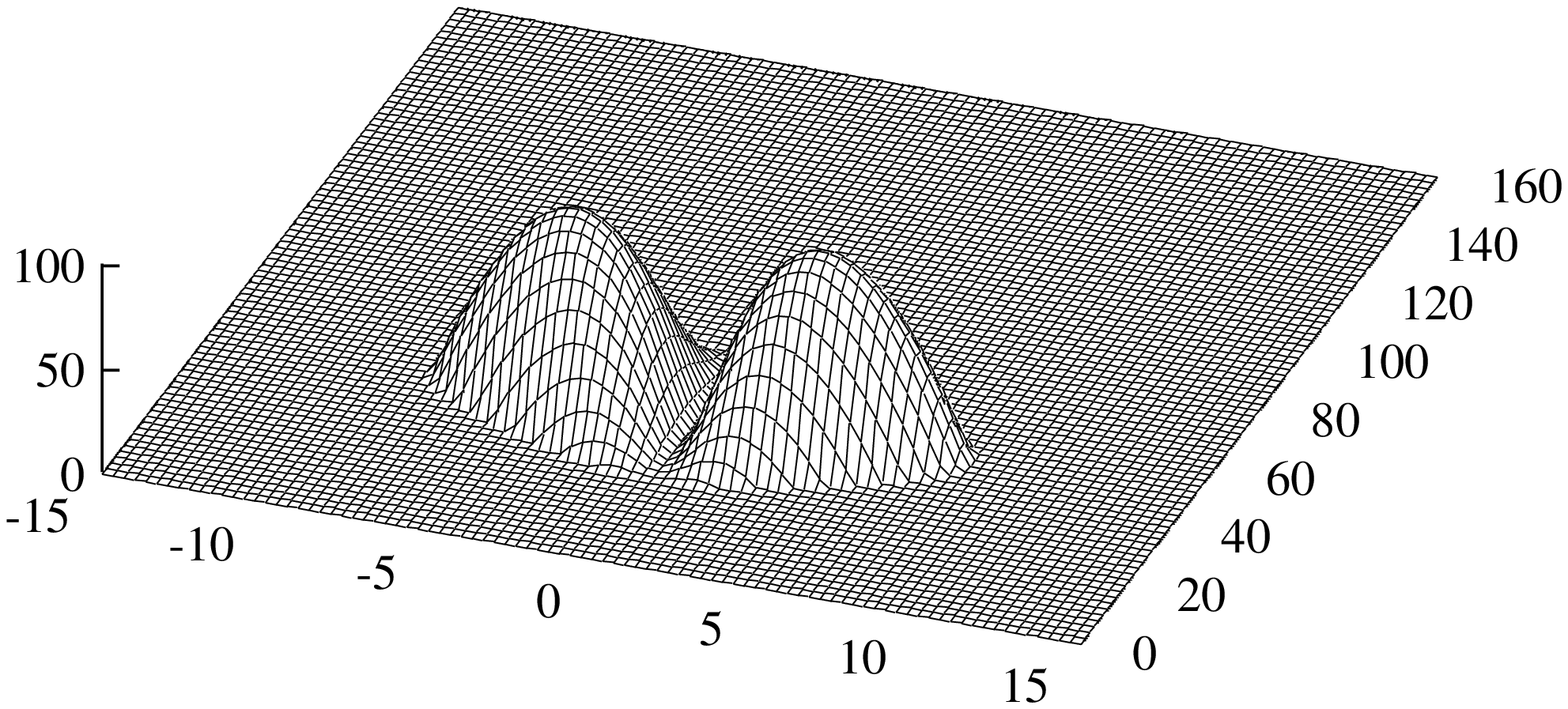}}
\put(1,5.8){\makebox(0,0){\large (a)}}
\put(1,4.5){\makebox(0,0){\large $\lambda(\rho_{\rm e},k)$ (1/h)}}
\put(7.5,-0.6){\makebox(0,0){\large $k$ (1/km)}}
\put(13,2.4){\makebox(0,0){\large $\rho_{\rm e}$}}
\put(13,1.8){\makebox(0,0){\large (veh/km)}}
\end{picture}
\begin{picture}(14,6.8)(0,-0.8)
\put(0,6){\epsfig{height=14cm, angle=-90, 
      bbllx=9cm, bblly=4.5cm, bburx=18.5cm, bbury=25cm, clip=,
      file=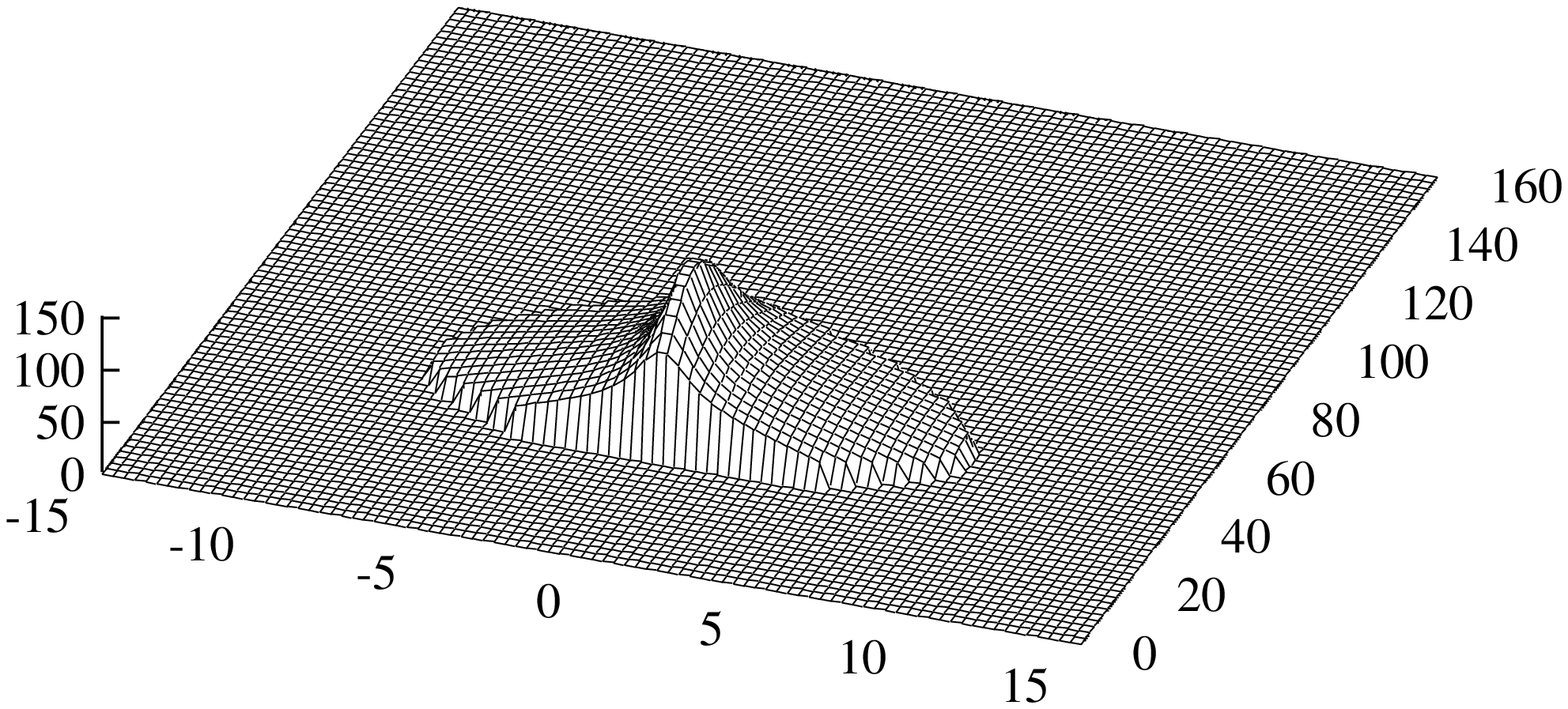}}
\put(1,5.8){\makebox(0,0){\large (b)}}
\put(1,4.7){\makebox(0,0){\large $-c(\rho_{\rm e},k)$}}
\put(1,4){\makebox(0,0){\large (km/h)}}
\put(7.5,-0.6){\makebox(0,0){\large $k$ (1/km)}}
\put(13,2.4){\makebox(0,0){\large $\rho_{\rm e}$}}
\put(13,1.8){\makebox(0,0){\large (veh/km)}}
\end{picture}
\end{center}
\caption[]{\protect (a) Illustration of the largest growth parameter 
$\lambda$ for the instability region (i.e.\ where $\lambda(\rho_{\rm e},k) \ge
0$). According to the K\"uhne-Kerner-Konh\"auser model (with $\tau = 11$\,s,
$\eta_0 = 436$\,km/h, $\sqrt{\Theta_0} = 54$\,km/h, and the equilbrium
velocity $V_{\rm e}(\rho)$ depicted in Fig.\ \protect\ref{eqvel}) traffic
flow is unstable at moderate densities and small absolute wavelengths $|k|$.
At $k=0$ we have marginal stability (i.e.\ $\lambda(\rho_{\rm e},0) = 0$)
due to the conservation of the number of vehicles. Instabilities at large
absolute wave numbers $|k|$ (i.e.\ small wave lengths) are surpressed by the
smoothing effect of viscosity. Although the model describes the instability
region qualitatively right, practical experience indicates that stop-and-go
traffic also develops at much higher densities.\protect\\
(b) The representation of the negative group velocity of the unstable
mode relative to $V_{\rm e}$ shows that emerging 
stop-and-go waves move in backward direction,
as expected. However, the relative propagation speeds seem to be somewhat
too large.} %Improved results for a refined traffic model are depicted
%in Fig.\ \protect\ref{fig2}.}}
\label{kernkon}
\end{figure}
The propagation speed of small perturbations relative to 
$V_{\rm e}(\rho_{\rm e})$ is given by the
relative {\em group velocity} 
\begin{equation}
 c(\rho_{\rm e},k) := \frac{\partial}{\partial k}
 \tilde{\omega}(\rho_{\rm e},k) = \pm \frac{\partial}{\partial k}
 \sqrt{\frac{1}{2}\left( \sqrt{\Re^2 + \Im^2} - \Re \right) } 
\end{equation}
(cf.\ Fig.\ \ref{kernkon}b).
At the transition from stability to instability we find
\begin{equation}
 c(\rho_{\rm e},k) = \pm \frac{1}{k} \sqrt{C_{\rm r}}
 = \pm \sqrt{\frac{\partial {\cal P}_{\rm e}}{\partial \rho}} \, .
\end{equation}
This formula is analogous to the one for the velocity of sound
in ordinary compressible fluids.
However, for traffic flows, the plus sign corresponds to the stable mode 
and the minus sign to the unstable mode, so that the forming stop-and-go waves
move in backward direction. This is in agreement with empirical
findings and solves a problem raised by Daganzo \cite{Dag}.
\item Helbing \cite{Hel,Gas} introduced a further dynamic equation
for the variance,
\begin{equation}
 \frac{\partial \Theta}{\partial t} + V \frac{\partial \Theta}{\partial r}
 = - \frac{2{\cal P}}{\rho} \frac{\partial V}{\partial r}
 - \frac{1}{\rho}\frac{\partial {\cal J}}{\partial r}
 + \frac{2}{\tau} (\Theta_{\rm e} - \Theta) \, ,
\end{equation} 
which bases on a gas-kinetic traffic model. This equation is analogous to 
the equation of {\em heat conduction}. However, it contains an additional
relaxation term $2(\Theta_{\rm e} - \Theta)/\tau$. Moreover, despite their
theoretical relationship, the variance $\Theta$
is better not denoted as temperature, and ${\cal J}$ is
not a heat flow but the {\em flux density of velocity variance}. 
\par
Another modification suggested by Helbing intended to take into account the
finite space requirements $s(V)$ by vehicles. In analogy to the
pressure relation of van der Waals for a gas of hard spheres
(cf. Eq.\ (\ref{Waals})), he
proposed the relation
\begin{equation}
 {\cal P} = \frac{\rho\Theta}{1 - \rho s(V)} \, .
% - \eta \frac{\partial V}{\partial r} \, ,
\end{equation}
%where ${\cal P}_{\rm id} = \rho \Theta - \eta_0 \partial V/\partial r$
%is the idealized pressure relation in the limit of small densities.
However, in Section \ref{hightraf} it will turn out that this relation cannot
simply be transfered from the theory of ordinary gases to the theory of
traffic flow: The dissipative and anisotropic interactions of vehicles
as well as their velocity-dependent space requirements
change the structure of the fluid-dynamic traffic equations considerably. 
\end{itemize}

\section{Paveri-Fontana's gas-kinetic traffic model} \label{PaFo}

Prigogine and coworkers \cite{Prig} were the first who proposed the
derivation of macroscopic traffic equations from a gas-kinetic
level of description. However, Paveri-Fontana \cite{Pav} noticed some
strange properties of their approach and suggested a modified model which 
will now be discussed. This model describes the temporal evolution of
the {\em phase-space density} $\hat{\rho}(r,v,v_0,t)$ of vehicles with
{\em desired velocity} $v_0$ and {\em (actual) velocity} $v$ at
place $r$ and time $t$, which is governed by the {\em continuity equation}
\begin{equation}
 \frac{\partial \hat{\rho}}{\partial t}
 + \frac{\partial}{\partial r} (\hat{\rho}v) + \frac{\partial}{\partial v}
 \left( \hat{\rho} \frac{dv}{dt} \right) + \frac{\partial}{\partial v_0}
 \left( \hat{\rho} \frac{dv_0}{dt} \right) = \left( \frac{\partial \hat{\rho}}
 {\partial t} \right)_{\rm dis} \, .
\label{gaskin}
\end{equation}
The {\em acceleration law} was specified by
\begin{equation}
 \frac{dv}{dt} = \frac{v_0 - v}{\tau} \, ,
\end{equation}
delineating an exponential adaptation of the actual velocity $v$ to the
individual desired velocity $v_0$ with a density-dependent relaxation time
$\tau(\rho)$. Since the individual desired velocity $v_0$ is usually assumed 
to be constant, we have 
\begin{equation}
 \frac{dv_0}{dt} = 0 \, .
\end{equation}
\par
According to Eq.\ (\ref{gaskin}), the substantial time derivative of the
phase-space density is given by temporal changes $(\partial
\hat{\rho}/\partial t)_{\rm dis}$ due to {\em discontinuously} modelled
velocity changes. Usually, {\em interaction processes} are delineated
by this term, since they happen %go off 
much faster than the acceleration 
processes \cite{Note1}. Paveri-Fontana proposed to use the
Boltzmann-like equation
\begin{eqnarray}
 \left( \frac{\partial \hat{\rho}} {\partial t} \right)_{\rm dis}
 &=& (1-p) \int\limits_{w>v} \!\! dw \int dw_0 \, |w-v|
 \hat{\rho}(r,v,w_0,t) \hat{\rho}(r,w,v_0,t) \nonumber \\
 &-& (1-p) \int\limits_{w<v} \!\! dw \int dw_0 \, |v-w|
 \hat{\rho}(r,w,w_0,t) \hat{\rho}(r,v,v_0,t) \, ,
\label{int1}
\end{eqnarray}
assuming that a slower vehicle can be immediately overtaken with
probability $p(\rho)$, and that the faster vehicle exactly decelerates to the
velocity of the slower one, if this is not possible. Whereas the first term
delineates an increase of $\hat{\rho}(r,v,v_0,t)$ due to vehicles with
velocity $w>v$ that must decelerate to velocity $v$, the second term
describes a decrease of $\hat{\rho}(r,v,v_0,t)$ due to vehicles with
velocity $v$ which must decelerate to a velocity $w<v$. 
According to (\ref{int1}), the interaction frequency of vehicles is 
proportional to their relative velocity $|v-w|$ and to the phase-space
densities $\hat{\rho}$ of the interacting vehicles. Their desired velocities
$v_0$ and $w_0$ are not changed by interactions.
\par
The gas-kinetic equation allows a derivation of macroscopic traffic equations
for the spatial vehicle {\em density}
\begin{equation}
 \rho(r,t) := \int dv \int dv_0 \, \hat{\rho}(r,v,v_0,t) 
\end{equation}
and the {\em velocity moments}
\begin{equation}
 \langle v^k (v_0)^l \rangle := \int dv \int dv_0 \, v^k (v_0)^l
 \frac{\hat{\rho}(r,v,v_0,t)}{\rho(r,t)} \, ,
\end{equation}
in particular the {\em mean velocity}
\begin{equation}
 V(r,t) := \langle v \rangle = \int dv \, v P(v;r,t) 
\end{equation}
and the velocity {\em variance}
\begin{equation}
 \Theta(r,t) := \langle (v-V)^2 \rangle 
 = \int dv \, (v-V)^2 P(v;r,t) \, ,
\end{equation}
where we have introduced the {\em velocity distribution}
\begin{equation}
 P(v;r,t) := \int dv_0 \, \frac{\hat{\rho}(r,v,v_0,t)}{\rho(r,t)}.
\end{equation}
The macroscopic equations are obtained by 
multiplication of the gas-kinetic equation (\ref{gaskin})
with $\psi(v):=1$, $v$, $v^2$ and subsequent integration
over $v$ and $v_0$. Finally, one finds the fluid-dynamic traffic equations
\begin{eqnarray}
 \frac{\partial \rho}{\partial t} + V \frac{\partial \rho}{\partial r}
 &=& - \rho \frac{\partial V}{\partial r} \, , \label{densder} \\
 \frac{\partial V}{\partial t} + V \frac{\partial V}{\partial r}
 &=& - \frac{1}{\rho} \frac{\partial {\cal P}}{\partial r}
 + \frac{1}{\tau} (V_{\rm e} - V) \, , \label{velder} \\
 \frac{\partial \Theta}{\partial t} + V \frac{\partial \Theta}{\partial r}
 &=& -\frac{2{\cal P}}{\rho} \frac{\partial V}{\partial r}
 - \frac{1}{\rho} \frac{\partial {\cal J}}{\partial r}
 + \frac{2}{\tau} (\Theta_{\rm e} - \Theta ) \, , 
\label{varder}
\end{eqnarray}
which was shown in Refs.\ \cite{Gas,Pav}. We recognize that the density
equation (\ref{densder}) and the velocity equation (\ref{velder}) 
have again the form of Eqs.\ (\ref{conteq}) and (\ref{veleq}). 
However, we have additionally found theoretical relations for the
{\em traffic pressure}
\begin{equation}
 {\cal P} := \rho \langle (v-V)^2 \rangle = \rho \Theta
\end{equation}
and the {\em equilibrium velocity}
\begin{equation}
 V_{\rm e} := V_0 - \tau(\rho) [1 - p(\rho)] \rho \Theta \, .
\label{velrel}
\end{equation}
Moreover, we have obtained the {\em variance equation} (\ref{varder})
with the {\em flux density of velocity variance}
\begin{equation}
 {\cal J} := \rho \langle (v-V)^3 \rangle \, ,
\end{equation}
the {\em equilibrium variance}
\begin{equation}
 \Theta_{\rm e} := {\cal C} - \frac{1}{2} \tau(\rho)[1-p(\rho)]
 \rho \langle (v-V)^3 \rangle \, ,
\end{equation}
and the {\em covariance}
\begin{equation}
 {\cal C} := \langle (v-V)(v_0 - V_0) \rangle \, .
\end{equation}
\par
For the evaluation of ${\cal J}(r,t)$ we need a mathematical expression 
for the {\em velocity distribution} $P(v;r,t)$.
This can be approximately obtained
by Grad's method of moments \cite{Grad}, which uses the expansion
\begin{equation}
 P(v;r,t) \approx \sum_{n=0}^N a_n(r,t) \frac{\partial^n}{\partial v^n}
 \frac{1}{\sqrt{2\pi\Theta}} \exp\left[ - \frac{(v-V)^2}{2\Theta} \right] \, .
\end{equation} 
The functions $a_n(r,t)$ can be expressed by the velocity moments 
$\langle v^k \rangle$ ($0\le k \le N$) which are governed by the 
macroscopic equations. In gas theory the macroscopic
equations are closed after the variance equation ($N=2$), since higher
velocity moments are varying on much faster time scales, so that their
adiabatic elimination is justified. For vehicular traffic as for granular
materials, a separation of time scales is not this simple. 
Therefore, one usually restricts to the
number $N$ of macroscopic equations, which allow to delineate
the instabilities under consideration. Sela and Goldhirsch \cite{grangas}
have shown that this method yields reliable and ``universal''
results in the sense that the inclusion of additional macroscopic equations 
gives no fundamental but only minor corrections. 
\par
From our discussion in Sec.\ \ref{models} we know that the continuity and
velocity equation are sufficient for a description of emerging
stop-and-go traffic. Therefore, we choose $N=1$, so that we must
express $\Theta(r,t)$ in dependence of $\rho(r,t)$ and $V(r,t)$:
\begin{equation}
  \Theta(r,t) := \Theta_{\rm e}(\rho(r,t),V(r,t)) \, .
\label{Neq1}
\end{equation} 
Because of the conditions $\langle 1 \rangle \stackrel{!}{=} 1$ and
$\langle v \rangle \stackrel{!}{=} V(r,t)$, we find $a_0(r,t) = 1$
and $a_1(r,t) = 0$, so that we get the Gaussian distribution
\begin{equation}
 P(v;r,t) \approx  \frac{1}{\sqrt{2\pi\Theta_{\rm e}}} 
 \exp\left[ - \frac{(v-V)^2}{2\Theta_{\rm e}} \right] \, .
\label{gauss}
\end{equation} 
This leads to \cite{Gas}
\begin{equation}
{\cal J} = 0 \qquad \mbox{and} \qquad \Theta_{\rm e} = {\cal C}_{\rm e} \, , 
\label{noskew}
\end{equation}
where ${\cal C}_{\rm e}$ is the {\em equilibrium covariance}
(cf.\ Fig.\ \ref{skew}).
In Ref.\ \cite{Gas} it has been shown that ${\cal C}_e$ is given by
the implicit equation
\begin{equation}
 {\cal C}_{\rm e} = \langle (v_0 - V_0)^2 \rangle - 
 2 \tau(\rho) [1-p(\rho)]  \rho {\cal C}_{\rm e} \sqrt{\frac{\Theta_{\rm e}}
 {\pi}} \, .
\label{eqcov}
\end{equation}
Therefore, $\Theta_{\rm e}$ is only a function of $\rho$ and not of
$V$ (cf.\ Fig.\ \ref{empvar}): 
Eqs.\ (\ref{noskew}) and (\ref{eqcov}) together with (\ref{velrel})
imply the relation
\begin{equation}
  \sqrt{\Theta_{\rm e}(\rho)} = - \frac{V_0 - V_{\rm e}(\rho)}{\sqrt{\pi}}
 + \sqrt{\frac{[V_0 - V_{\rm e}(\rho)]^2}{\pi} + \langle (v_0 - V_0)^2
 \rangle } \, .
\label{eqvar}
\end{equation}
In homogeneous traffic situations, the {\em variance of desired velocities}
$\Theta_0 := \langle (v_0 - V_0)^2 \rangle$ should be a constant and, in
particular, independent of density.
Therefore, relation (\ref{eqvar}) predicts that the variance is still
finite when the equilibrium velocity $V_{\rm e}$ vanishes. This is, of
course, a paradoxial result. 
\par
\begin{figure}[htbp]
\unitlength1cm
\begin{center}
\begin{picture}(14,10.6)(0,-0.8)
\put(0,9.8){\epsfig{height=14cm, angle=-90, 
      bbllx=50pt, bblly=50pt, bburx=554pt, bbury=770pt, % clip=,
      file=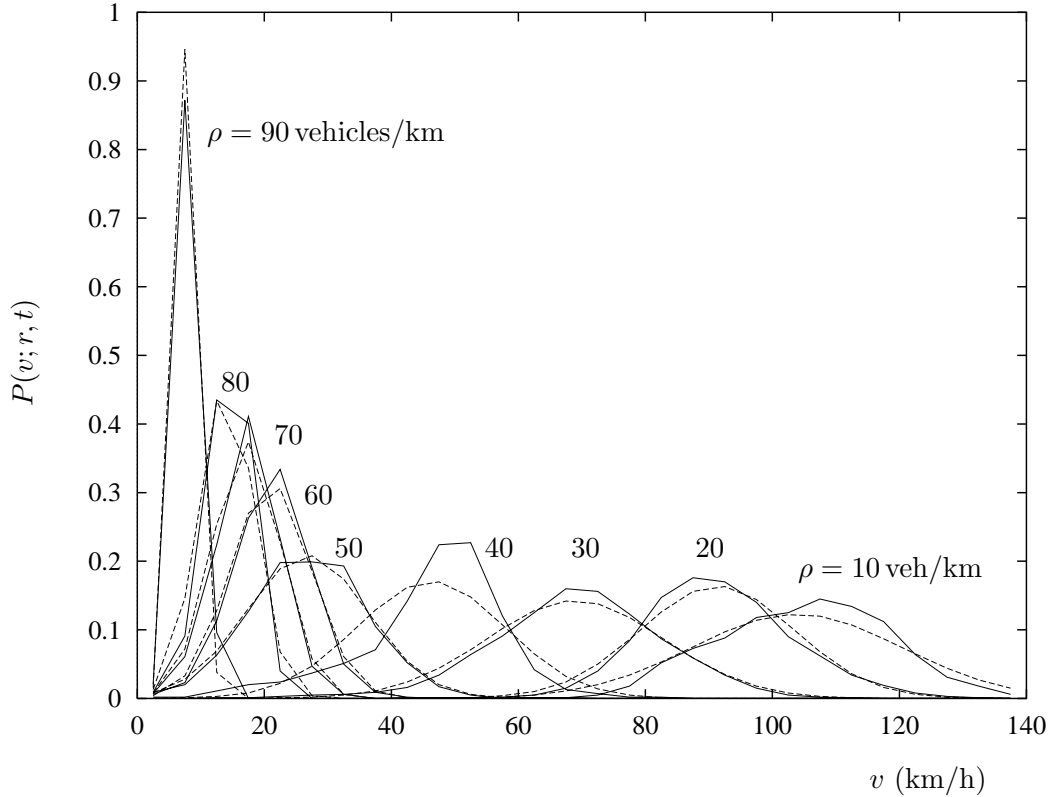}}
\put(12.2,-0.6){\makebox(0,0){\small $v$ (km/h)}}
\put(0.2,5.1){\makebox(0,0){\rotate[l]{\hbox{\small $P(v;r,t)$}}}}
\put(4.2,8){\makebox(0,0){\small $\rho=90$\,vehicles/km}}
\put(3,4.7){\makebox(0,0){\small 80}}
\put(3.7,4){\makebox(0,0){\small 70}}
\put(4.1,3.2){\makebox(0,0){\small 60}}
\put(4.5,2.5){\makebox(0,0){\small 50}}
\put(6.5,2.5){\makebox(0,0){\small 40}}
\put(7.65,2.5){\makebox(0,0){\small 30}}
\put(9.3,2.5){\makebox(0,0){\small 20}}
\put(11.7,2.2){\makebox(0,0){\small $\rho=10$\,veh/km}}
\end{picture}
\end{center}
\caption[]{Comparison of empirical velocity distributions at different
densities (---) with frequency polygons of 
grouped Gaussian velocity distributions with
the same mean value and variance (--~--). The class interval lengths
are 5\,km/h. A signigicant deviation of the empirical relations from the
respective discrete Gaussian approximations is only found at density
$\rho = 40$\,vehicles/km, where the two minute averages of the single
vehicle data may have been too long due to rapid stop-and-go waves
(cf.\ the mysterious ``knee'' at $\rho \approx 40$\,veh/km 
in Fig.~\protect\ref{empvar}). 
The velocity distributions keep their unimodal form even at high densities.
(Data: Dutch highway A9 with constant speed limit 120\,km/h.)}
\label{v_vert}
\end{figure}
\begin{figure}[htbp]
\unitlength1cm
\begin{center}
\begin{picture}(14,10.6)(0,-0.8)
\put(0,9.8){\epsfig{height=14cm, angle=-90, 
      bbllx=50pt, bblly=50pt, bburx=554pt, bbury=770pt, % clip=,
      file=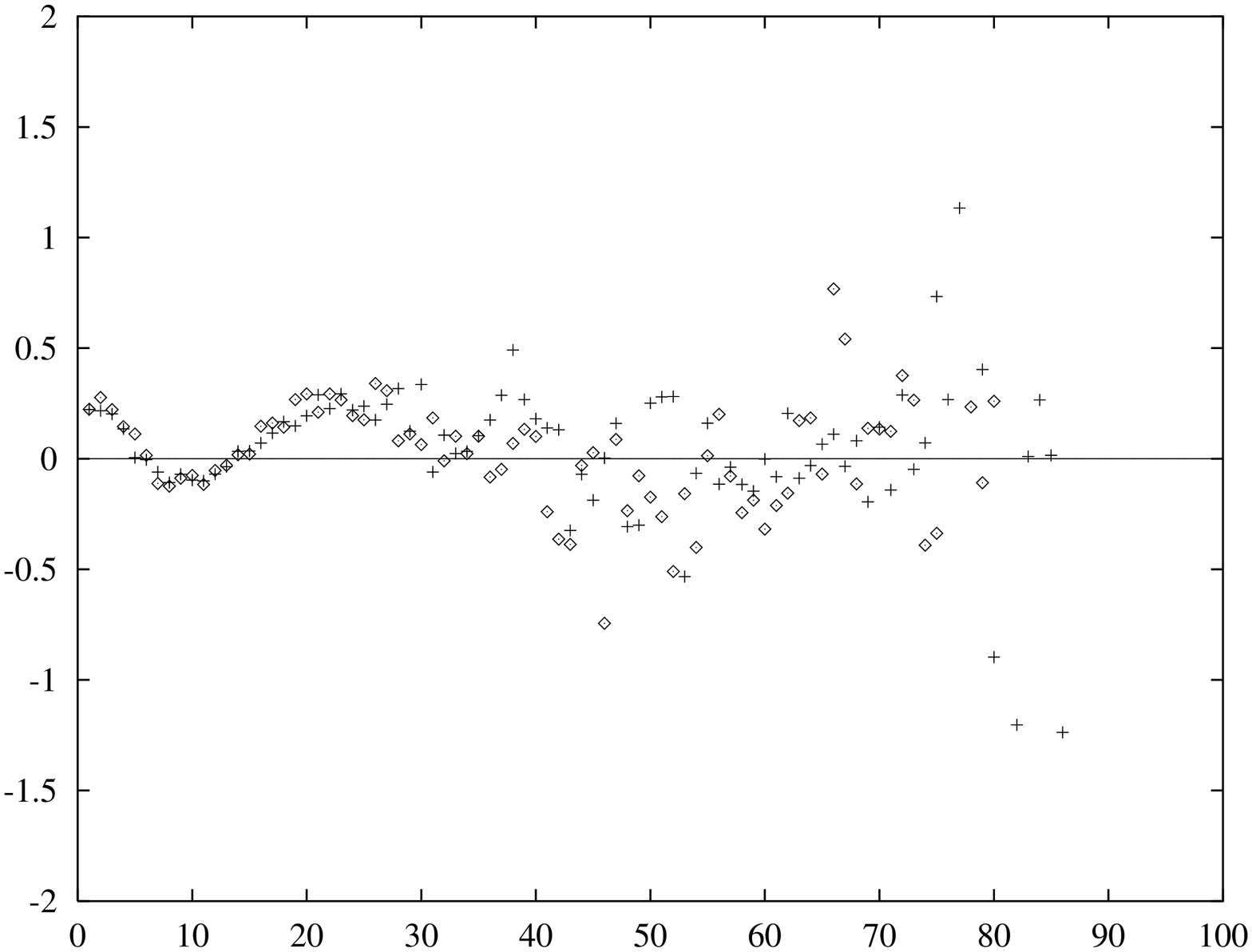}}
\put(12,-0.6){\makebox(0,0){\small $\rho$ (vehicles/km)}}
\put(0.2,5.1){\makebox(0,0){\rotate[l]{\hbox{\small $\displaystyle 
 \frac{\langle (v-V)^3 \rangle}{\Theta^{3/2}}$}}}}
\end{picture}
\end{center}
\caption[]{Empirical skewness of the velocity distribution on the basis of
mean values of one minute averages ($\Diamond$: October 14, 1994; +:
November 2, 1994). The absolute value of the skewness is rather small
and mostly lies between 0 and 0.5.
%A systematical deviation from 0 is only found for
%densities 10\,veh/km $\le \rho\le$ 30\,veh/km. 
(The variation
of the data at higher densities results from the small amount of one minute
data that could be averaged.)}
\label{skew}
\end{figure}
Another problem was recognized by Shvetsov \cite{Priv}:
According to Paveri-Fontana's equation, vehicles are not decelerated 
to velocities less than the minimum desired velocity of all drivers.
Therefore, it cannot describe %the typical situation of 
the development of so-called {\em phantom traffic jams,} where all drivers
want to drive fast, but produce %get stuck in 
a slowly moving traffic jam. 
\par
As a consequence, we must modify Paveri-Fontana's equation somewhat. 
This is done in the next two sections.
\begin{figure}[htbp]
\unitlength1cm
\begin{center}
\begin{picture}(14,10.6)(0,-0.8)
\put(0,9.8){\epsfig{height=14cm, angle=-90, 
      bbllx=50pt, bblly=50pt, bburx=554pt, bbury=770pt, % clip=,
      file=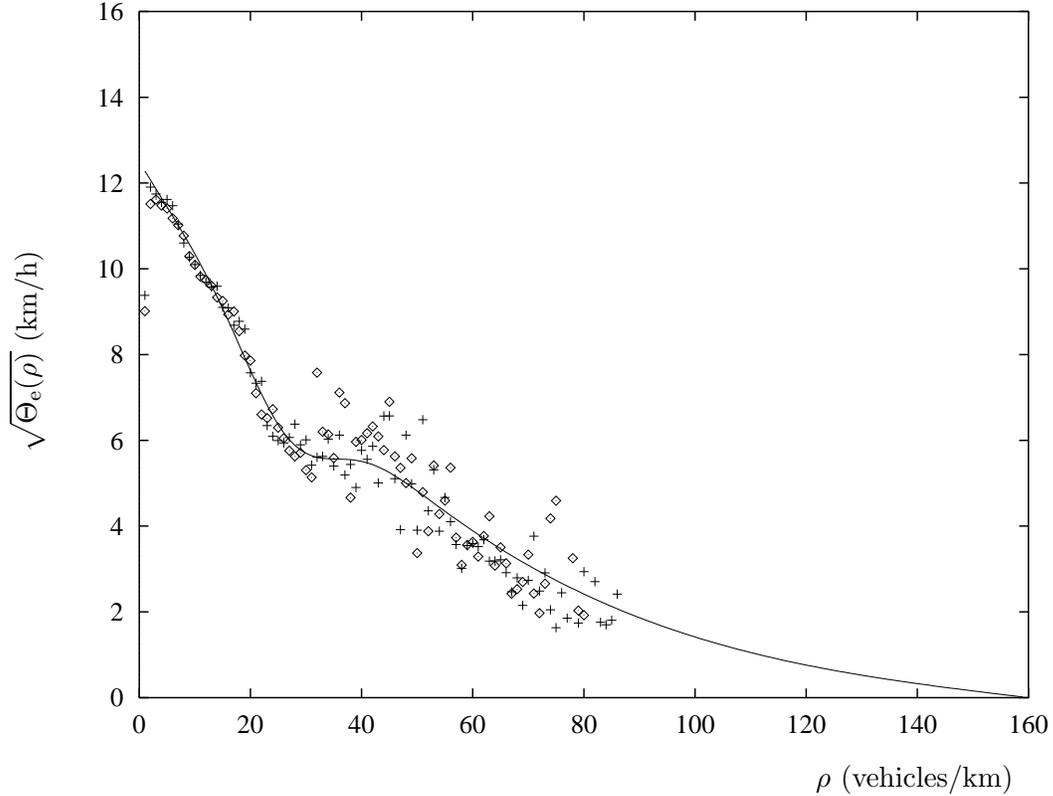}}
\put(12,-0.6){\makebox(0,0){\small $\rho$ (vehicles/km)}}
\put(0.2,5.1){\makebox(0,0){\rotate[l]{\hbox{\small 
 $\sqrt{\Theta_{\rm e}(\rho)}$ (km/h)}}}}
\end{picture}
\end{center}
\caption[]{Illustration of the empirical equilibrium variance-density relation
$\Theta_{\rm e}(\rho)$,
evaluated like in Fig.\ \protect\ref{eqvel} ($\Diamond$: October 14, 1994;
+: November 2, 1994; ---: fit function).}
\label{empvar}
\end{figure}

\section{Imperfect driving} \label{imp}

Up to now, we have assumed perfectly driving vehicles that are able to
keep a constant speed and to exactly adapt to the velocity of a
slower vehicle in front. Therefore, we will replace Paveri-Fontana's
interaction term by the more general formula
\begin{equation}
 \left( \frac{\partial \hat{\rho}}{\partial t} \right)_{\rm dis}
 := \left( \frac{\partial \hat{\rho}}{\partial t} \right)_{\rm diff}
 + \left( \frac{\partial \hat{\rho}}{\partial t} \right)_{\rm int} \, ,
\end{equation}
where 
\begin{equation}
 \left( \frac{\partial \hat{\rho}}{\partial t} \right)_{\rm diff}
 := \frac{1}{2} \frac{\partial^2}{\partial v^2} ({\cal D} \hat{\rho})
\end{equation}
describes {\em velocity fluctuations} due to imperfect speed control.
The diffusion function ${\cal D}$ must be chosen in such a way that 
the velocity-diffusion term $(\partial \hat{\rho}/\partial t)_{\rm diff}$
cannot produce negative velocities. In the following, we will assume
\begin{equation}
 D(v) = \frac{2\alpha v^2}{\tau} 
\label{drel}
\end{equation}
in order to obtain a dimensionless relation $\alpha(\rho)$. According to
Eq.\ (\ref{drel}), velocity fluctuations are related to 
the time scale $\tau$ of velocity adaptation. Moreover, 
comparing (\ref{drel}) with the formula $\langle |\Delta x|^2
\rangle  = {\cal D} \Delta t$ for spatial diffusion with diffusion constant 
${\cal D}$, the absolute velocity displacement $|\Delta v|$ is 
approximately {\em proportional} to $v$ during small time intervals $\Delta t$,
which sounds plausible. 
\par
Now, we will generalize the interaction term. By
\begin{eqnarray}
 \left( \frac{\partial \hat{\rho}} {\partial t} \right)_{\rm int}
 &=& \int dw \!\!\int\limits_{v'>w} \!\!  dv' \!\!\int\limits_{w' \ge v} \!\!
 dw' \int dw_0 \, |w-v'|
 \sigma(w',v|w,v')\hat{\rho}(r,w,w_0,t) \hat{\rho}(r,v',v_0,t) \nonumber \\
 &-& \int\limits_{w<v} \!\! dw \int dv' \!\!\int\limits_{w'\ge v'}\!\! 
 dw' \int dw_0 \, |v-w|
 \sigma(w',v'|w,v) \hat{\rho}(r,w,w_0,t) \hat{\rho}(r,v,v_0,t) \qquad
\label{int2}
\end{eqnarray}
we can take into account vehicles that do not exactly adapt to the
velocity of a slower car in front. $\sigma(w',v'|w,v)$
denotes the {\em differential cross section} and describes the probability
with which two vehicles have velocities $v'$ and $w'$ after their interaction,
if they had velocities $v$ and $w$ before. In the following, we will choose
\begin{equation}
 \sigma(w',v'|w,v) := 
 \frac{1-p}{\beta w} \exp \left( - \frac{w - v'}{\beta w}
 \right) \delta (w' - w) 
\label{sig}
\end{equation}
where $\delta(.)$ denotes Dirac's delta function. According to this formula,
the velocity $w$ of the slower vehicle is not influenced by the velocity 
of a faster vehicle with velocity $v$ behind it (i.e.\ $w' = w$). 
The exponential function reflects that the faster vehicle slows down to 
a velocity $v'\le w$. Although the velocity $v'=w$ is most likely,
smaller velocities $v' < w$ also occur with a certain probability due to
an imperfect adaptation of velocity. For the degree $\beta$ of imperfection
we have $0 \le \beta \ll 1$. In the limit $\beta = 0$, formula (\ref{sig})
becomes $\sigma(w',v'|w,v) = (1-p) \delta(v' - w)\delta(w'-w)$, so that
we get back Paveri-Fontana's interaction term (\ref{int1}).
\par
In order to obtain the macroscopic traffic equations corresponding to the
generalized gas-kinetic equation, we have to evaluate the interaction
terms
\begin{eqnarray}
 {\cal I}(\psi) &:=& \int dv \int dv_0 \, \psi(v)
 \left( \frac{\partial \hat{\rho}} {\partial t} \right)_{\rm int} \nonumber \\
 &=& \rho^2(r,t) \int dv \int dw \!\!\int\limits_{v'>w} \!\!  dv' \!\!
 \int\limits_{w' \ge v}\!\! dw' \, 
 \psi(v) |w-v'| \sigma(w',v|w,v')P(w;r,t) P(v';r,t) \nonumber \\
 &-& \rho^2(r,t) \int dv \!\!\int\limits_{w<v} \!\! dw \int dv' \!\!
 \int\limits_{w'\ge v'}\!\! dw' \, 
 \psi(v) |v-w| \sigma(w',v'|w,v) P(w;r,t) P(v;r,t) \nonumber \\
 &=& \rho^2(r,t) \int dv \!\!\int\limits_{w<v} \!\! dw \int dv' \!\!
 \int\limits_{w' \ge v'} \!\! dw' \,
 |v-w| \sigma(w',v'|w,v) \nonumber \\
 & & \times [\psi(v') - \psi(v)] P(w;r,t) P(v;r,t) \, ,
\label{vehint}
\end{eqnarray}
where we have interchanged variables to obtain the final result. With
(\ref{gauss}) and (\ref{sig}) we find ${\cal I}(1) = 0$ and
\begin{eqnarray}
 {\cal I}(v) &=& - (1-p) \rho^2 \left[ \Theta_{\rm e} 
 \left( 1 - \frac{\beta}{2} \right) 
 + \beta V \sqrt{\frac{\Theta_{\rm e}}{\pi}} \right] \, , \nonumber \\
 {\cal I}((v-V)^2) &\approx & (1-p) \rho^2 \left( \beta V \Theta_{\rm e}
 + 2 \beta^2 V^2 \sqrt{\frac{\Theta_{\rm e}}{\pi}} \right) \, .
\end{eqnarray}
Since the diffusion term $(\partial \hat{\rho}/\partial t)_{\rm diff}$
yields no contribution to the density and velocity equation, but
the contribution $2\alpha (V^2 + \Theta_{\rm e})/\tau$ to the variance 
equation, we arrive at the corrected equilibrium relations
\begin{eqnarray}
 V_{\rm e}(\rho,V) &=& V_0 - \tau (1-p) \rho \left[ \Theta_{\rm e} 
 \left( 1 - \frac{\beta}{2} \right) 
 + \beta V \sqrt{\frac{\Theta_{\rm e}}{\pi}} \right] \, , 
\label{eqrel1} \\
 \Theta_{\rm e}(\rho,V) &=& {\cal C}_{\rm e} + \alpha (V^2 + \Theta_{\rm e})
 + \frac{\tau}{2} (1-p) \rho \left( \beta V \Theta_{\rm e}
 + 2 \beta^2 V^2 \sqrt{\frac{\Theta_{\rm e}}{\pi}} \right) \, .
\label{eqrel2}
\end{eqnarray} 
Note that the equilibrium relations $V_{\rm e}$ and $\Theta_{\rm e}$ not
only depend on the density $\rho$, but also on the mean velocity
$V$, now. The previous formulas result for $\alpha = \beta = 0$.

\subsection{Determination of the parameters from empirical data}

The calibration of the functions $\alpha(\rho)$, $\beta(\rho)$,
$p(\rho)$, and $\tau(\rho)$ is a difficult problem, since we only have
empirical relations for $V_{\rm e}(\rho)$ and $\Theta_{\rm e}(\rho)$
(cf.\ Figs.\ \ref{eqvel} and \ref{empvar}).
Therefore, we set $\beta = 0$ in order to reduce the number of parameters.
(Setting $\alpha = 0$ does not allow to describe the finite equilibrium
variance at small densities if ${\cal C}_{\rm e} = 0$, see below.)
Since the equilibrium covariance ${\cal C}_{\rm e}$ is not anymore
given by relation (\ref{eqcov}), we calculate it via Eq.\ (\ref{eqrel2}): 
\begin{equation}
 {\cal C}_{\rm e}(\rho) = (1-\alpha) \Theta_{\rm e}(\rho,V_{\rm e}(\rho)) 
 - \alpha [V_{\rm e}(\rho)]^2 \, .
\end{equation}
This expression is completely determined by 
$V_{\rm e}(\rho)$ and $\Theta_{\rm e}(\rho)$, if we define
\begin{equation}
 \alpha := \lim_{\rho\rightarrow \rho_{\rm max}}
 \frac{\Theta_{\rm e}(\rho,V_{\rm e}(\rho))}{[V_{\rm e}(\rho)]^2 
 + \Theta_{\rm e}(\rho,V_{\rm e}(\rho))} \, .
\end{equation}
Consequently, the equilibrium variance 
\begin{equation}
  \Theta_{\rm e}(\rho,V) = \frac{{\cal C}_{\rm e}(\rho) + \alpha V^2}
 {1 - \alpha}
\label{thetarel}
\end{equation}
is given by the diffusion effect $\alpha V^2/(1-\alpha)$ at high densities,
whereas it is dominated by the covariance ${\cal C}_{\rm e}$ at low
densities, which is very plausible. 
In the case of speed limits, however, we
have ${\cal C}_{\rm e} \approx 0$, since all vehicles have approximately
the same desired velocity $v_0 \approx V_0$. In this situation we define
\begin{equation}
 \alpha(\rho) := \frac{\Theta_{\rm e}(\rho,V_{\rm e}(\rho))}
 {[V_{\rm e}(\rho)]^2 + \Theta_{\rm e}(\rho,V_{\rm e}(\rho))} \, .
\end{equation}
\par
\begin{figure}[htbp]
\unitlength1cm
\begin{center}
\begin{picture}(13.2,10.6)(0.8,-0.8)
\put(0,9.8){\epsfig{height=14cm, angle=-90, 
      bbllx=50pt, bblly=50pt, bburx=554pt, bbury=770pt, % clip=,
      file=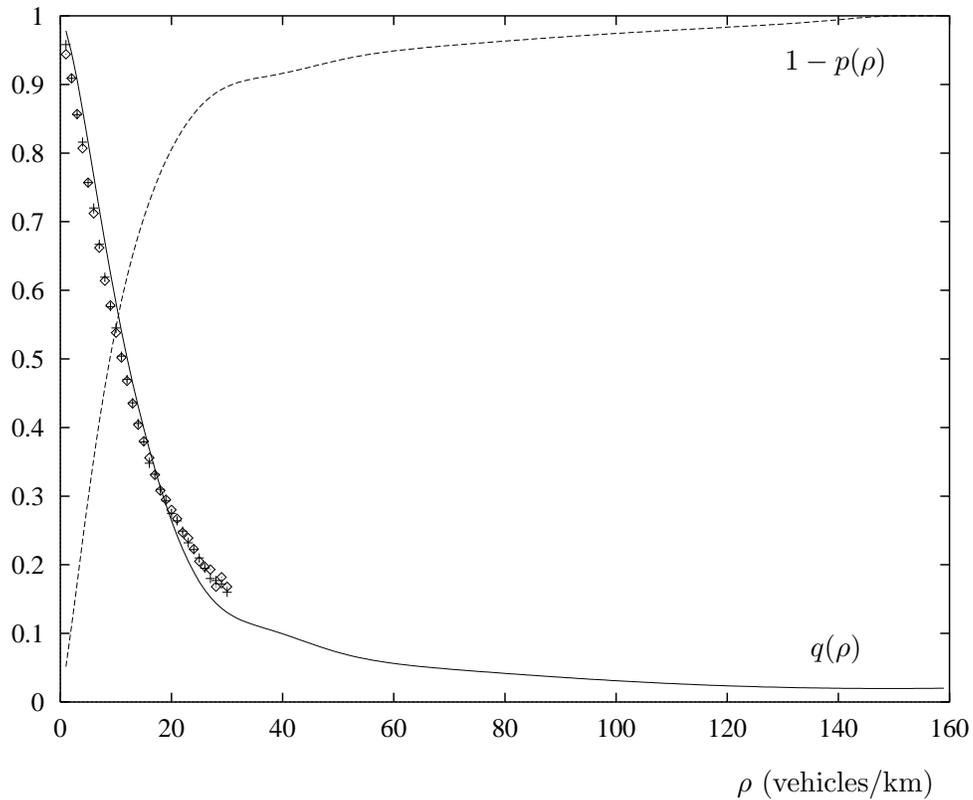}}
\put(12,-0.6){\makebox(0,0){\small $\rho$ (vehicles/km)}}
\put(12,9){\makebox(0,0){\small $1-p(\rho)$}}
\put(12,1.2){\makebox(0,0){\small $q(\rho)$}}
\end{picture}
\end{center}
\caption[]{Illustration of different relations estimated from empirical
traffic data: The density-dependent probability $(1-p)$ that a fast car
cannot immediately overtake a slower one (--~--) starts at zero and reaches
the
value one in the limit $\rho \rightarrow \rho_{\rm max}$. The proportion
$q(\rho)$ of freely moving vehicles $q(\rho)$ is about one for very small
densities and approaches a finite value with growing density (---).
Symbols represent the proportion of vehicles that have a time headway
greater than $2.5T$, where $T= 0.8$\,s corresponds to the safe
time headway ($\Diamond$: October 14, 1994; +: November 2, 1994).}
\label{taup}
\end{figure}

The product $\tau(\rho)[1-p(\rho)]$ can be easily obtained via
Eq.\ (\ref{velab}), which is a generalization of Eq.\ (\ref{eqrel1}):
\begin{equation}
  \tau(\rho) [1-p(\rho)] = \frac{V_0 - V_{\rm e}(\rho)}
 {\rho \chi \Theta_{\rm e}(\rho,V_{\rm e}(\rho))} \, .
\end{equation}
A determination of the factors $p(\rho)$ and $\tau(\rho)$ is only possible
by means of additional assumptions. A detailed analysis shows that
the relaxation time is given by the relation 
\begin{equation}
\tau(\rho) = \frac{q(\rho)}{\tau_0} \, ,
\end{equation}
where $q(\rho)$ is the {\em proportion of freely moving vehicles} and
$\tau_0 \approx 8$\,s the relaxation time of vehicles which
are not impeded during their acceleration \cite{Hab}. In addition, a 
complicated relation of the form $p(\rho,q(\rho),V_{\rm e}(\rho))$ can be
derived (cf.\ Ref.\ \cite{Hab}). The resulting density-dependent functions
$q(\rho) = \tau_0/\tau(\rho)$ and $[1-p(\rho)]$ are depicted in 
Figure \ref{taup}. It is very important that, although $p(\rho)$ approaches 
the value 0 in the limit $\rho \rightarrow \rho_{\rm max}$, the relaxation time
$\tau(\rho)$ remains finite, since the maximum density $\rho_{\rm max}$ is
somewhat less than the inverve of the {\em average vehicle length} $l_0$.

\section{Dynamics at high densities} \label{high}

Up to now, vehicles have been implicitly treated as point-like objects.
Therefore, the model must now be corrected for the finite space requirements
of vehicles. This will lead to further modifications of the interaction
term. In order to illustrate the applied method, we will first discuss
the gas-kinetic and fluid-dynamic description of granular material
like sand, powder, or pills \cite{densgas,grangas,granfluids}.

\subsection{Granular flow} \label{highgran}

Granular materials have found a broad interest due to their various instability
phenomena like density waves, avalanches, cluster or heap formation,
convection, or size segregation \cite{overview,phenomena}. 
In order to work out the similarities
with traffic flow, we will focus on the description of density waves in sand 
falling through a narrow pipe \cite{pipewaves}. Similar to the discussion by
Rieth\-m\"uller et al. \cite{grangas} we assume a vertical tube of diameter
$s$, in which spherical grains (or cylinders of height $s$) with
radius $s/2$ are falling. Since grains do not have a desired velocity $v_0$,
we are confronted with the {\em phase-space density} $\tilde{\rho}(r,v,t)$ of 
grains with velocity $v$, this time. The corresponding gas-kinetic equation
is
\begin{equation}
 \frac{\partial \tilde{\rho}}{\partial t} +  \frac{\partial }{\partial r}
 (\tilde{\rho} v) + \frac{\partial}{\partial v} \left(
 \tilde{\rho} \frac{dv}{dt} \right) 
 = \frac{1}{2} \frac{\partial^2}{\partial v^2} ({\cal D}\tilde{\rho})
 + \left(\frac{\partial \tilde{\rho}}{\partial t}\right)_{\rm int} \, .
\label{grankin}
\end{equation}
It exactly corresponds to the gas-kinetic traffic equation 
in the case of a speed limit $V_0$, where the desired velocities
of all drivers approximately agree ($v_0 \approx V_0$), so that $v_0$ 
is not a variable anymore, but a fixed parameter. Since
%(i.e. $\tilde{\rho}(r,v,t) = \hat{\rho}(r,v,V_0,t)\delta(v_0-V_0)$). 
velocity changes of grains result from {\em acceleration due to gravity}
$g$ as well as from sliding friction at the wall (and by displacement of
air) with a {\em friction coefficient} $\gamma$, we have the acceleration
law
\begin{equation}
  \frac{dv}{dt} = g - \gamma v \, .
\end{equation}
The diffusion term $\frac{1}{2}{\cal D}\partial^2\tilde{\rho}/\partial v^2$ 
with {\em diffusion constant} ${\cal D} := 2 \gamma \Theta_0$ describes
variations of the grain velocities $v$ due to fluctuating influences
of the wall (and the displaced air) in accordance with the {\em
  fluctuation-dissipation theorem}. 
\par
Multiplying the gas-kinetic equation (\ref{grankin}) with $\psi(v):=1$,
$v$, or $v^2$, and integrating over $v$ gives the macroscopic equations
\begin{eqnarray}
 \frac{\partial \rho}{\partial t} + V \frac{\partial \rho}{\partial r}
 &=& - \rho \frac{\partial V}{\partial r} \, , \label{densgran} \\
 \frac{\partial V}{\partial t} + V \frac{\partial V}{\partial r}
 &=& - \frac{1}{\rho} \frac{\partial {\cal P}}{\partial r}
 + ( g  - \gamma V ) + \frac{{\cal I}(v)}{\rho} \, , \label{velgran} \\
 \frac{\partial \Theta}{\partial t} + V \frac{\partial \Theta}{\partial r}
 &=& -\frac{2{\cal P}}{\rho} \frac{\partial V}{\partial r}
 - \frac{1}{\rho} \frac{\partial {\cal J}}{\partial r}
 + 2\gamma (\Theta_0 - \Theta ) + \frac{{\cal I}((v-V)^2)}{\rho} \, , 
\label{vargran}
\end{eqnarray}
where we have taken into account the conservation of the number of vehicles
(${\cal I}(1) = 0$). A comparison with the fluid-dynamic traffic equations
shows that the equations agree with each other if
we substitute $g \leftrightarrow V_0/\tau$, $\gamma \leftrightarrow 1/\tau$,
$\Theta_0 \leftrightarrow \alpha (V^2+\Theta)$, and specify the
interaction term like in Eq.\ (\ref{int1}) with $v_0 \equiv V_0 \equiv
w_0$. Although this specification was
suggested by Rieth\-m\"uller et al. \cite{grangas}, 
we will apply the theoretical relations for
dense gases and granular materials \cite{densgas}, instead, 
in order to include effects due to high densities and momentum 
conservation. The interaction terms have then the form
\begin{eqnarray}
 {\cal I}(\psi) &=& \int dv \!\!\int\limits_{w<v} \!\! dw \int dv'
 \!\!\int\limits_{w' \ge v'} \!\! dw' \, |v-w| \sigma(w',v'|w,v) 
 [\psi(v') - \psi(v)] \tilde{\rho}_2(r+s,w;r,v;t) \nonumber \\
 &+& \int dv \!\!\int\limits_{w>v} \!\! dw \int dv'
 \!\!\int\limits_{w'\le v'}\!\! dw' \, |v-w| \sigma(v',w'|v,w) 
 [\psi(v') - \psi(v)] \tilde{\rho}_2(r,v;r-s,w;t) \, . \nonumber \\
 & & \label{genint}
\end{eqnarray}
In contrast to the formula (\ref{vehint}) for vehicular interactions, we have
an additional contribution due to backward interactions of grains
(last term). Moreover, we have taken into account that pushed grains
are located at places $r+s$ or $r-s$. Finally,
the {\em pair distribution} function $\tilde{\rho}_2$ describes
velocity-correlations between interacting grains. 
\par
The expression (\ref{genint}) can be simplified by interchanging
variables:
\begin{eqnarray}
{\cal I}(\psi)
 &=& \int dv \!\!\int\limits_{w<v} \!\! dw \int dv'
 \!\!\int\limits_{w' \ge v'} \!\! dw' \, |v-w| \sigma(w',v'|w,v) \nonumber \\
 & & \times \{ [\psi(v') - \psi(v)] \tilde{\rho}_2(r+s,w;r,v;t) 
 \vphantom{\int} \nonumber \\
 & & \,\, + [\psi(w') - \psi(w)] \tilde{\rho}_2(r,w;r-s,v;t) \} \, .
\label{psiI}
\end{eqnarray}
Applying the Taylor approximation
\begin{equation}
 \tilde{\rho}_2(r+s,w;r,v;t) \approx \tilde{\rho}_2(r,w;r-s,v;t)
 + s \frac{\partial}{\partial r} \tilde{\rho}_2(r+s,w;r,v;t) 
\end{equation}
yields
\begin{equation}
 {\cal I}(\psi) = {\cal I}_{\rm s}(\psi) - \frac{\partial {\cal I}_{\rm f}
 (\psi)} {\partial r}
\end{equation}
with the {\em source-like contribution}
\begin{eqnarray}
 {\cal I}_{\rm s}(\psi) &=& 
\int dv \!\!\int\limits_{w<v} \!\! dw \int dv'
 \!\!\int\limits_{w' \ge v'} \!\! dw' \, |v-w| \sigma(w',v'|w,v) \nonumber \\
 & & \times \{ [\psi(v') +\psi(w')] - [\psi(v) + \psi(w)]\} 
 \tilde{\rho}_2(r,w;r-s,v;t)
\end{eqnarray}
and the {\em flux-like contribution}
\begin{equation}
{\cal I}_{\rm f}(\psi)
 = - s  \int dv \!\!\int\limits_{w<v} \!\! dw \int dv'
 \!\!\int\limits_{w'\ge v'} dw' \, |v-w| \sigma(w',v'|w,v) 
 [\psi(v') - \psi(v)] \tilde{\rho}_2(r+s,w;r,v;t) 
\end{equation}
describing collisional transfer. In this representation we immediately
see that the source-like contributions vanish for {\em collisional
invariants} $\psi$, for which $\psi(v') + \psi(w') = \psi(v) + \psi(w)$
holds. As a consequence, we have only flux-like contributions to
the velocity and variance equation for ordinary gases or fluids. 
However, in granular material the conservation of kinetic energy gets lost
due to inelastic, dissipative interactions. Therefore, a source-like 
contribution is expected in the equation for the so-called 
{\em granular temperature} $\Theta$ (cf.\ Eq.\ (\ref{granvar})). 
Since momentum is conserved during
granular collisions, these yield only a flux-like contribution to the
velocity equation, which results in a corrected pressure relation
(cf.\ Eq.\ (\ref{Waals})).
Therefore, we have again a velocity equation of the form (\ref{veleq})
with a density-{\em in}dependent equilibrium velocity $V_{\rm e} = g/\gamma$.
Consequently, the instability condition for the granular flow
is given by Eq.\ (\ref{instcond}), if we assume the validity of approximation 
(\ref{Neq1}). Due to $\partial V_{\rm e}/\partial \rho = 0$, the
stationary and homogeneous solution of the granular
density and velocity equation is stable at all densities $\rho_{\rm e}$.
For this reason, the approximation (\ref{Neq1}) is not valid for granular
flows, and we must apply Grad's method with $N\ge 2$. For $N=2$ we obtain
the Gaussian-shaped Maxwell-Boltzmann distribution
\begin{equation}
 P(v;r,t) \approx  \frac{1}{\sqrt{2\pi\Theta}} 
 \exp\left[ - \frac{(v-V)^2}{2\Theta} \right] \, ,
\end{equation} 
because $\langle 1 \rangle \stackrel{!}{=} 1$, 
$\langle v \rangle \stackrel{!}{=} V$, and
$\langle (v-V)^2 \rangle \stackrel{!}{=} \Theta$ imply 
$a_0(r,t) = 1$ and $a_1(r,t) = a_2(r,t) = 0$.  
\par
The explicit calculation of the interaction terms ${\cal I}(\psi)$
now calls for a specification of the differential cross section 
$\sigma(w',v'|w,v)$ of granular collisions. Since both interacting
grains have to be equivalently treated (isotropy condition), the collision law
$v' = \mu v + \mu' w$ implies $w' = \mu w + \mu' v$. Due to momentum
conservation $(v'+w' = v+w)$ we must set $\mu' = (1-\mu)$. Therefore, we 
find the relation
\begin{equation}
 \sigma(w',v'|w,v) = \delta(v' - [\mu v + (1-\mu) w ] )
 \delta(w' - [\mu w + (1-\mu) v ] ) \, ,
\end{equation}
which is invariant with respect to interchanging the particles. $\mu$ is a
characteristic parameter of the granular material. It is related to
the amount of {\em energy dissipation} because of
\begin{equation}
 (v^{\prime\,2} + w^{\prime\,2}) = (v^2+w^2)
- 2 \mu (1-\mu)(v-w)^2 \, . 
\end{equation}
Since collisional energy must not be produced during the interactions,
we find $0 \le \mu \le 1$. Moreover,
since we have $v>w$ before the collision
and $v' \le w'$ after the collision, we obtain the further restriction
$\mu \le 1/2$. For $\mu = 0$ we have completely
elastic collisions with an interchange of particle velocities, so that
the interaction terms in the macroscopic equations (\ref{velgran}) and
(\ref{vargran}) vanish in the limit $s \ll 1/\rho$
(like for ordinary gases or fluids). However,
due to its analogy with vehicular traffic, we will focus 
on the extremely inelastic case $\mu = 1/2$ in which both particles 
have the same velocity after their collision.
\par
Next, we will specify the pair distribution function $\tilde{\rho}_2$ of
interacting grains. This is
usually expressed by their one-particle phase-space densities $\tilde{\rho}$
in the following way \cite{densgas}:
\begin{equation}
 \tilde{\rho}_2(r+s,w;r,v;t) = \chi(r+s/2,t) \tilde{\rho}(r+s,w,t)
 \tilde{\rho}(r,v,t) \, .
\end{equation}
The factor 
\begin{equation}
 \chi(r+s/2,t) := \frac{1}{1- \rho(r+s/2,t) s}
\end{equation}
reflects the increase of the particle interaction rate \cite{intfreq}
because the
grains are extended by an amount $s/2$ around their centers, so that
they collide earlier. A first-order Taylor approximation 
of $\tilde{\rho}_2$ gives
\begin{eqnarray}
 & & \chi(r\pm s/2,t) \tilde{\rho}(r\pm s,w,t) \tilde{\rho}(r,v,t)
 \approx \chi(r,t) \tilde{\rho}(r,w,t) \tilde{\rho}(r,v,t) 
 \vphantom{\sum_b} \nonumber \\
 & & \qquad \times \left\{ 1 \pm \frac{s}{2\chi} 
 \frac{\partial \chi}{\partial r} 
 \pm s \left[ \frac{1}{\rho} \frac{\partial \rho}{\partial r}
 + \frac{v-V}{\Theta} \frac{\partial V}{\partial r}
 + \frac{1}{2\Theta} \left( \frac{(v-V)^2}{\Theta} - 1\right) 
 \frac{\partial \Theta} {\partial r} \right] \right\} \, .
\end{eqnarray} 
With this, evaluating the interaction terms ${\cal I}(\psi)$ and 
neglecting higher order derivatives as well as products of derivatives
(i.e.\ neglecting Navier-Stokes and Burnett corrections) finally yields the
continuity equation and the following Euler-like equations for granular flows:
\begin{equation}
 \frac{\partial V}{\partial t} + V \frac{\partial V}{\partial r}
 = - \frac{1}{\rho}\frac{\partial}{\partial r} ({\cal P} +
 {\cal P}_{\rm corr}) + (g-\gamma V) \, ,
\label{granvel}
\end{equation}
\begin{equation}
 \frac{\partial \Theta}{\partial t} + V \frac{\partial \Theta}{\partial r}
 = - \frac{2}{\rho} [ {\cal P} + (1+3\mu){\cal P}_{\rm corr}]
 \frac{\partial V}{\partial r} + 2\gamma (\Theta_0 - \Theta)
 - \mu (1-\mu) \frac{8}{\sqrt{\pi}}
 \frac{\rho\Theta^{3/2}}{1 - \rho s} \, .
\label{granvar}
\end{equation}
That is, in the velocity equation (\ref{granvel})
the interaction term yields an additional contribution
\begin{equation}
 {\cal P}_{\rm corr} = (1-\mu) {\cal P} \frac{\rho s}{1-\rho s}
\end{equation}
to the pressure. This diverges in the limit
$\rho \rightarrow \rho_{\rm max} = 1/s$ of extreme densities,
whereas ${\cal P} = \rho \Theta$ vanishes,
since the equilibrium variance $\Theta_{\rm e}(\rho)$ given by
the implicit equation
\begin{equation}
 \Theta_{\rm e}(\rho) = \Theta_0 - \mu (1-\mu) \frac{4}{\gamma\sqrt{\pi}}
 \frac{\rho\Theta_{\rm e}^{3/2}(\rho)}{1 - \rho s}
\end{equation}
vanishes. In the elastic case $\mu = 0$ we obtain the
formula of van der Waals for the {\em total pressure}
${\cal P}_{\rm tot}$ in a gas of hard spheres:
\begin{equation}
 {\cal P}_{\rm tot} := {\cal P} + {\cal P}_{\rm corr}
 = {\cal P} \left( 1 + (1-\mu) \frac{\rho s}{1 - \rho s} \right)
 \stackrel{\mu=0}{=} \frac{{\cal P}}{1 - \rho s} \, .
\label{Waals}
\end{equation}
The last term in the variance equation (\ref{granvar})
results from energy dissipation during granular collisions. Moreover, we have
discovered a new contribution $3\mu{\cal P}_{\rm corr}$ to the
pressure, which (to the knowledge of the author)
has not been reported elsewhere, presumably because it only 
plays a role in very inelastic cases ($\mu \not\approx 0$).
\par
\begin{figure}[htbp]
\unitlength1cm
\begin{center}
\begin{picture}(14,12)(0,-0.8)
\put(0,8){\epsfig{height=14cm, angle=-90, 
      bbllx=5cm, bblly=4.3cm, bburx=18.5cm, bbury=25cm, %clip=,
      file=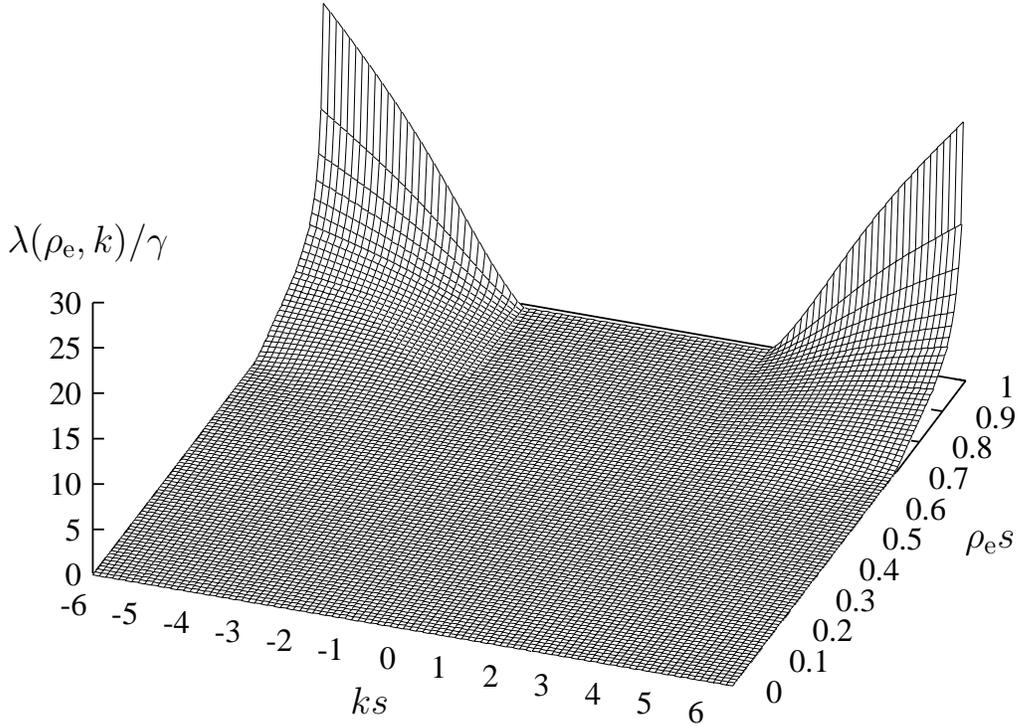}}
\put(1,5.5){\makebox(0,0){\large $\lambda(\rho_{\rm e},k)/\gamma$}}
\put(4.75,-0.6){\makebox(0,0){\large $ks$}}
\put(13,1.5){\makebox(0,0){\large $\rho_{\rm e}s$}}
\end{picture}
\end{center}
\caption{Illustration of the largest growth parameter $\lambda$ in units of
$\gamma$ for the instability region (i.e.\ where $\lambda(\rho_{\rm e},k)/ \gamma
\ge 0$). The wave numbers in units of $1/s$
are restricted to the range $-2\pi \le ks \le 2\pi$,
since wave lengths shorter than the diameter $s$ of the grains cannot be
propagated by the granular material.
Granular flow is stable for wave numbers $k \approx 0$. Emergent density
waves appear above a density-dependent 
critical value of $|k|$. These develop faster with growing 
absolute wave numbers $|k|$ and increasing densities $\rho_{\rm e}$.
%Varying the degree of dissipation $\mu$ (here: $\mu = 1/2$) causes only
%quantitative but not qualitative changes of the instability diagram. 
A smaller degree of dissipation $\mu$ (here: $\mu = 1/2$)
results in an expansion of the instability region towards
higher densities $\rho_{\rm e}$ and smaller absolute wave numbers $|k|$.
Decreasing the noise level $\Theta_0$ leads to a reduction of the instability
region, until it vanishes below a certain critical value.}
\label{fig1}
\end{figure}

A linear stability analysis about the stationary and spatially homogeneous
solution of Eqs.\ (\ref{densgran}), (\ref{granvel}), and (\ref{granvar})
leads to the characteristic polynomial
\begin{eqnarray}
 & & - \tilde{\lambda}^3 + \tilde{\lambda}^2 \left( 
 2 \frac{\partial \Theta_{\rm
     e}}{\partial \Theta} - 3 \right) \gamma + \tilde{\lambda}
 \left\{ 2 \gamma^2 \left( \frac{\partial \Theta_{\rm e}}{\partial \Theta}
 - 1 \right) \right. \nonumber \\
 & & \quad - k^2 \left. \left[ \frac{\partial {\cal P}_{\rm tot}}{\partial \rho}
 + \frac{2}{(\rho_{\rm e})^2} ( {\cal P}_{\rm tot} + 3 \mu {\cal P}_{\rm
   corr}) \frac{\partial {\cal P}_{\rm tot}}{\partial \Theta}
  \right] \right\} \nonumber \\ 
 & & - 2 \gamma k^2 \left[ \frac{\partial {\cal P}_{\rm tot}}{\partial \rho}
 \left( 1 - \frac{\partial \Theta_{\rm e}}{\partial \Theta} \right)
 + \frac{\partial {\cal P}_{\rm tot}}{\partial \Theta} 
 \frac{\partial \Theta_{\rm e}}{\partial \rho} \right] \stackrel{!}{=} 0 \, .
\label{charpol}
\end{eqnarray}
Its numerical investigation shows that 
the Euler-like equations for granular flows are unstable if the 
{\em noise level} $\Theta_0$ is sufficiently large (cf.\ Fig.\ \ref{fig1}).
The mechanism of the clustering instability originates from the increase of
granular pressure with growing density. This causes a reduction of velocity
(cf.\ Eq.~(\ref{granvel})) which results in a further compression
(cf.\ Eq.~(\ref{densgran})). Since Eq.\ (\ref{charpol}) implies
\begin{equation}
 \tilde{\lambda}(\rho_{\rm e},k)  = \tilde{\lambda}(\rho_{\rm e},-k) 
 \qquad \mbox{and} \qquad 
 \tilde{\omega}(\rho_{\rm e},k)  = \tilde{\omega}(\rho_{\rm e},-k) \, ,
\end{equation}
the {\em relative propagation speed}
\begin{equation}
  u(\rho_{\rm e},k) := \frac{\tilde{\omega}(\rho_{\rm e},k)}{k}
 = -\frac{\tilde{\omega}(\rho_{\rm e},-k)}{-k} = -u(\rho_{\rm e},-k)
\end{equation}
and the {\em relative group velocity}
\begin{equation}
  c(\rho_{\rm e},k) := \frac{\partial \tilde{\omega}(\rho_{\rm e},k)}
 {\partial k} = - \frac{\partial \tilde{\omega}(\rho_{\rm e},-k)}
 {\partial (-k)} = - c(\rho_{\rm e},-k)
\end{equation}
of perturbations with respect to  $V_{\rm e}$ are {\em antisymmetric}
functions in $k$. This is in contrast to the situation for vehicular
traffic (cf.\ Fig.\ \ref{fig2}b).

\subsection{Traffic flow} \label{hightraf}

Our previously discussed traffic model is now very easily extended to
the description of dense traffic. Again, we can apply formula
(\ref{psiI}). However, inserting (\ref{sig}) shows that the last term,
which corresponds to backward interactions, vanishes. % this time.
Therefore, the only difference between (\ref{vehint}) and (\ref{psiI}) is
the replacement of $\rho^2 P(v;r,t)P(w;r,t)$ by the pair distribution function
$\tilde{\rho}_2(r+s,w;r,v;t)$. In addition, we must take into account
that vehicles require a {\em velocity}-dependent space of about
\begin{equation}
 s(V) := \frac{1}{\rho_{\rm max}} + TV \, ,
\end{equation}
since drivers keep on average a {\em save distance} of length
$TV$, where $T \approx 0.8$\,s is about the {\em reaction time}.
As a consequence, we get the modified relation
\begin{equation}
 \tilde{\rho}_2(r+s,w;r,v;t) = \chi(r+TV,t)\rho(r+s(V),t)
 P(w;r+s(V),t) \rho(r,t) P(v;r,t)
\end{equation}
with
\begin{equation}
 \chi(r+TV,t) = \frac{1}{1 - \rho(r+TV,t) s(V)} \, .
\end{equation}
$r+TV(r,t)$ is the {\em interaction point} of a vehicle at place $r$
with a vehicle at place $r+s(V(r,t))$. In contrast to our discussion of
grains, $r$ and $r+s(V)$ are not the centers but the fronts of the vehicles.
\par
Now, we can evaluate the collision terms (\ref{psiI}). Applying
Taylor approximations for $\chi(r+TV,t)$ and
$\tilde{\rho}(r+s(V),w,t)$, we finally obtain the velocity equation
\begin{equation}
 \frac{\partial V}{\partial t} + V \frac{\partial V}{\partial r}
 = - a_1 \frac{\partial \rho}{\partial r} 
 + a_2 \frac{\partial V}{\partial r} - a_3 \frac{\partial \Theta}{\partial r}
 - b_1 \frac{\partial^2 \rho}{\partial r^2}
 + b_2 \frac{\partial^2 V}{\partial r^2} 
 - b_3 \frac{\partial^2 \Theta}{\partial r^2}
 + \frac{1}{\tau} (V_{\rm e} - V ) \, .
\end{equation}
We have neglected products of partial 
derivatives (Burnett corrections), but
taken into account second order derivatives (Navier-Stokes terms).
The abbreviations used in the above equation are:
\begin{eqnarray}
 a_1 &=& \frac{\Theta}{\rho} + (1-p)\rho\left(\frac{s\chi}{\rho}  
 + TV \frac{\partial \chi}{\partial \rho} \right) \left[ \Theta \left( 1 -
 \frac{\beta}{2} \right) + \beta V \sqrt{\frac{\Theta}{\pi}} \right] \, , 
 \nonumber \\
 a_2 &=& (1-p) \rho \left\{ s \chi \left[ \left( 2 - \frac{3}{2} \beta
 \right) \sqrt{\frac{\Theta}{\pi}} + \frac{\beta}{2} V \right]
 - TV \frac{\partial \chi}{\partial V} \left[ \Theta 
 \left( 1 - \frac{\beta}{2} \right) + \beta V \sqrt{\frac{\Theta}{\pi}} \right]
   \right\} \, , \nonumber \\
 a_3 &=& 1 + (1-p) \rho s \chi \left( \frac{1-\beta}{2} 
 + \frac{\beta V}{4\sqrt{\pi \Theta}} \right) \nonumber \\
 b_1 &=& (1-p)\rho\left(\frac{s^2\chi}{2\rho} + \frac{(TV)^2}{2} 
 \frac{\partial \chi}{\partial \rho} \right) \left[ \Theta \left( 1 -
 \frac{\beta}{2} \right) + \beta V \sqrt{\frac{\Theta}{\pi}} \right] \, , 
 \nonumber \\
 b_2 &=& (1-p) \rho \left\{ \frac{s^2\chi}{2} 
 \left[ \left( 2 - \frac{3}{2} \beta 
 \right) \sqrt{\frac{\Theta}{\pi}} + \frac{\beta}{2} V \right]
 - \frac{(TV)^2}{2} \frac{\partial \chi}{\partial V} \left[ \Theta \left( 1 -
 \frac{\beta}{2} \right) + \beta V \sqrt{\frac{\Theta}{\pi}} \right] 
 \right\} \, , \nonumber \\
 b_3 &=& (1-p) \rho \frac{s^2\chi}{2} \left( \frac{1-\beta}{2} 
 + \frac{\beta V}{4\sqrt{\pi \Theta}} \right) \, ,
\label{abbr}
\end{eqnarray}
and
\begin{equation}
 V_{\rm e} = V_0 - (1-p) \tau \rho \chi 
 \left[ \Theta \left( 1 - \frac{\beta}{2} \right) + \beta V \sqrt{\frac{
 \Theta}{\pi}} \right] \, .
\label{velab}
\end{equation}
Again, we will eliminate the variance equation by means of the approximation
$\Theta(r,t) \approx \Theta_{\rm e}(\rho(r,t),V(r,t))$. In the case
${\cal C}_{\rm e} \approx 0$ related to a speed limit,
the {\em equilibrium 
variance} $\Theta_{\rm e}(\rho,V)$ is given by the implicit equation
\begin{equation}
\Theta_{\rm e}(\rho,V) = \alpha (V^2 + \Theta_{\rm e}) 
 + \frac{\tau}{2} (1-p)\rho \chi \left( \beta V \Theta_{\rm e}
 + 2 \beta^2 V^2 \sqrt{\frac{\Theta_{\rm e}}{\pi}} \right) \, ,
\end{equation}
which is obtained by evaluation of the variance equation.
Consequently, $\Theta_e$ vanishes when $V$ becomes zero, which is required
for consistency \cite{Hel,Gas}. Applying the above
approximation for the variance and again neglecting
products of partial derivatives (i.e. Burnett corrections), we find
\begin{equation}
 \frac{\partial V}{\partial t} + V \frac{\partial V}{\partial r}
 = - \frac{1}{\rho} \frac{\partial {\cal P}_{\rm tot}}{\partial \rho}
 \frac{\partial \rho}{\partial r} + a \frac{\partial V}{\partial r}
 - b \frac{\partial^2 \rho}{\partial r^2}
 + \frac{\eta}{\rho} \frac{\partial^2 V}{\partial r^2} 
 + \frac{1}{\tau} ( V_{\rm e} - V ) \, ,
\label{refeq}
\end{equation}
where 
\begin{eqnarray}
 \frac{\partial {\cal P}_{\rm tot}}{\partial \rho} 
 &:=& \rho \left( a_1 + a_3 \frac{\partial
 \Theta_{\rm e}}{\partial \rho}\right) \, , \nonumber \\
 \eta &:=& \rho \left( b_2 - b_3 \frac{\partial \Theta_{\rm e}}
 {\partial V} \right) \nonumber \\
 a &:=& a_2 - a_3 \frac{\partial \Theta_{\rm e}}{\partial V} 
  \, , \nonumber \\
 b &:=& b_1 + b_3 \frac{\partial \Theta_{\rm e}}{\partial \rho} \, ,
\end{eqnarray}
and $V_{\rm e}$ are functions of $\rho$ and $V$. 
\par
It is obvious that the loss of momentum conservation and the
velocity-dependence of the vehicular space requirements have drastically
changed the structure of the velocity equation: The terms containing $a$ and
$b$ are completely
new compared to the Navier-Stokes equation for ordinary or granular
fluids \cite{densgas} and compared to all previous macroscopic traffic models
(cf.\ Sec.\ \ref{models}). Fortunately, the model solves
the problems which were mentioned in Section \ref{intro}: 
\begin{itemize}
\item The variation of individual vehicle velocities is taken into account
by the variance $\Theta_{\rm e}(\rho,V)$.
\item The equilibrium velocity $V_{\rm e}$ and the equilibrium variance
$\Theta_{\rm e}$ are monotonously decreasing functions that vanish
at the finite density $\rho_{\rm max}$.
\item The equilibrium variance $\Theta_{\rm e}(\rho,V)$ vanishes when
$V(r,t)=0$.
\item In equilibrium
(i.e.\ for $\rho(r,t) = \rho_{\rm e}$ and $V(r,t) = V_{\rm e}(\rho_{\rm e})$)
the density-gradient $\partial {\cal P}/\partial \rho$ of the total 
traffic pressure ${\cal P}$ is non-negative, so that the latter
is a monotonously increasing function of $\rho_{\rm e}$ 
(cf.\ Fig.\ \ref{dpress}). The singularity of the pressure 
at $\rho = \rho_{\rm max}$ guarantees that the upper bound
$\rho_{\rm max} < \rho_{\rm bb}$ cannot be exceeded by the density $\rho(r,t)$.
\begin{figure}[htbp]
\unitlength1cm
\begin{center}
\begin{picture}(14,10.6)(0,-0.8)
\put(0,9.8){\epsfig{height=14cm, angle=-90, 
      bbllx=50pt, bblly=50pt, bburx=554pt, bbury=770pt, % clip=,
      file=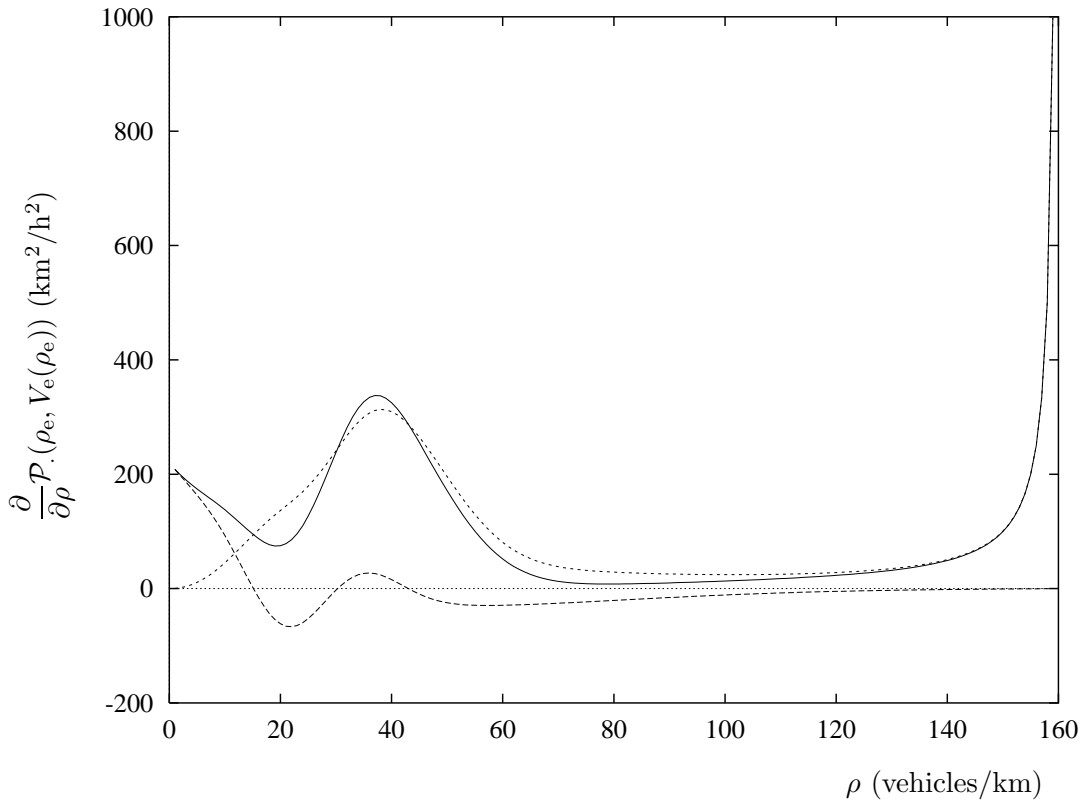}}
\put(12,-0.6){\makebox(0,0){\small $\rho$ (vehicles/km)}}
\put(0,5.1){\makebox(0,0){\rotate[l]{\hbox{\small $\displaystyle
\frac{\partial}{\partial 
\rho} {\cal P}_{.}(\rho_{\rm e},V_{\rm e}(\rho_{\rm e}))$ (km$^2$/h$^2$)}}}}
\end{picture}
\end{center}
\caption[]{Comparison of the density-gradients of the total traffic pressure
${\cal P}_{\rm tot}$ (---), the idealized pressure
${\cal P} = \rho \Theta$ of point-like objects (--~--),
and the correction term ${\cal P}_{\rm corr}$ (-~-~-), estimated from
empirical data. For $\rho \approx 20$\,veh/km and large densities,
the idealized traffic pressure ${\cal P}$ of point-like objects decreases with
growing density. However, this is more than compensated by the pressure
correction ${\cal P}_{\rm corr}$
due to the finite space requirements of vehicles. In particular,
the increase of pressure with growing density diverges in the limit
$\rho \rightarrow \rho_{\rm max}$.}
\label{dpress}
\end{figure}
\item We were able to explain viscosity as an effect of the finite
distance $s(V)$ between interacting vehicles. Moreover, we obtained
a theoretical expression for the viscosity $\eta(\rho,V)$, 
which not only depends on the density, but also on the mean velocity. 
In equilibrium $\rho(r,t) = \rho_{\rm e}$, $V(r,t) = V_{\rm e}(\rho_{\rm e})$,
the viscosity is non-negative. The singularity of viscosity at 
$\rho = \rho_{\rm max}$ causes that extreme changes of
$V(r,t)$ and $\rho(r,t)$ are smoothed out, so that the shock-formation
problem is solved.
\begin{figure}[htbp]
\unitlength1cm
\begin{center}
\begin{picture}(14,10.6)(0,-0.8)
\put(0,9.8){\epsfig{height=14cm, angle=-90, 
      bbllx=50pt, bblly=50pt, bburx=554pt, bbury=770pt, % clip=,
      file=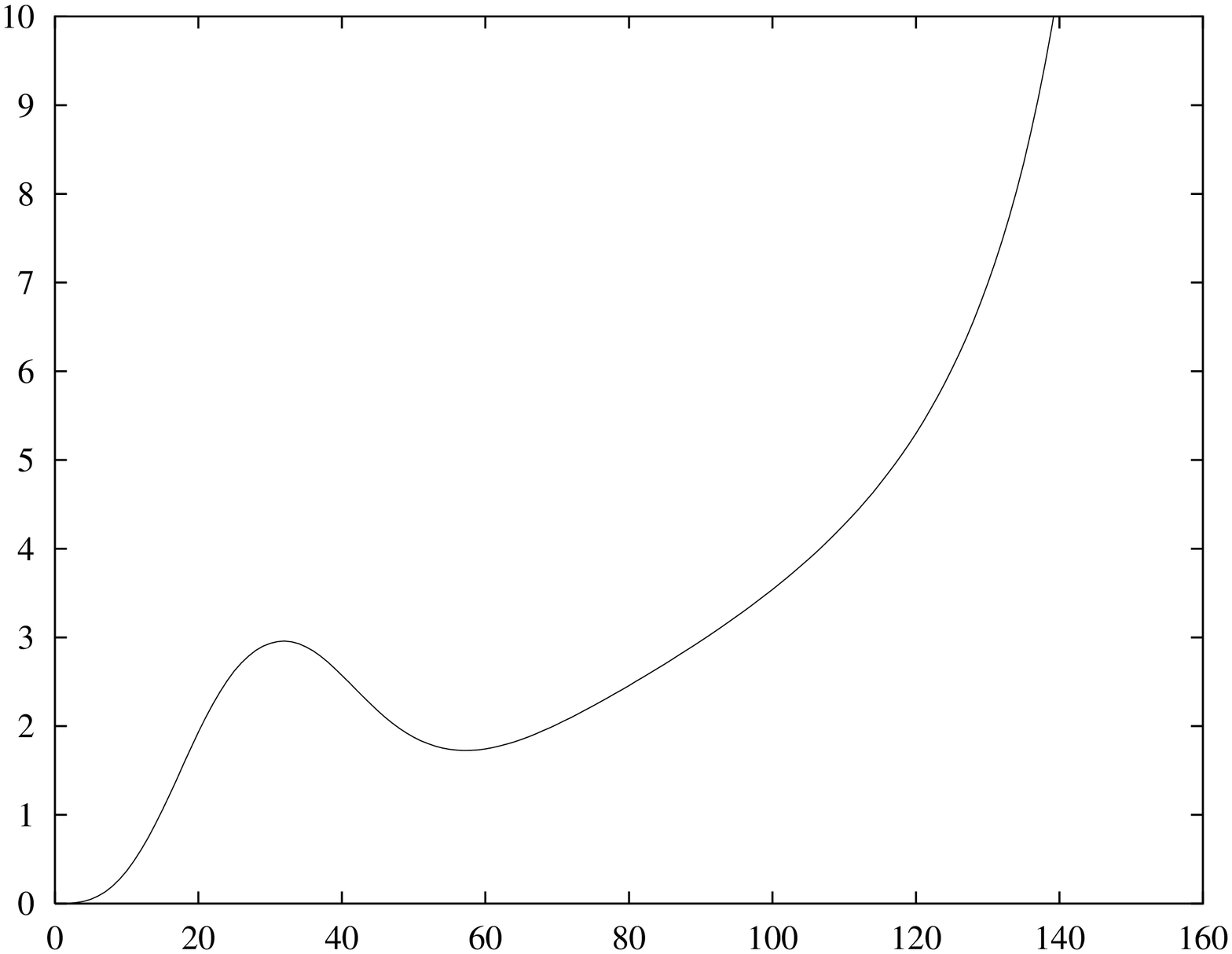}}
\put(12,-0.6){\makebox(0,0){\small $\rho$ (vehicles/km)}}
\put(0.2,5.1){\makebox(0,0){\rotate[l]{\hbox{\small $\eta(\rho_{\rm e},V_{\rm
          e}(\rho_{\rm e}))$ (km/h)}}}}
\end{picture}
\end{center}
\caption[]{Illustration of the viscosity function estimated from empirical
data. Obviously, the viscosity is non-negative, as expected, and diverges
at large densities. Viscosity causes the smoothing effect which is
necessary to avoid shock-like structures and to facilitate numerical 
simulations of the macroscopic traffic equations.}
\label{eta}
\end{figure}
%\item Non-negativity also holds for the functions $a(\rho,V)$ and $b(\rho,V)$.
%For this reason and since density $\rho(r,t)$
%increases where the mean velocity $V(r,t)$ increases (cf.\ Eq.\ (\ref{conteq})),
%the term containing $a$ has a similar effect like the pressure term
%and the term containing $b$ a similar effect like the viscosity term.
\item The fluid-dynamic traffic model (\ref{densder}), (\ref{refeq}) is able
to describe the emergence of stop-and-go traffic at medium densities
(cf.\ Fig.\ \ref{fig2}a). A linear stability analysis leads to the
characteristic polynomial
\begin{equation}
 \tilde{\lambda}^2 + \tilde{\lambda} \left( \frac{\eta k^2}{\rho_{\rm e}}
 + \frac{1}{\tau} - {\rm i} k a - \frac{1}{\tau} \frac{\partial V_{\rm e}}
 {\partial V} \right) + {\rm i} k \rho_{\rm e}
 \left( - \frac{{\rm i}k}{\rho_{\rm e}}
 \frac{\partial {\cal P}_{\rm tot}}{\partial \rho} 
 + \frac{1}{\tau}\frac{\partial
 V_{\rm e}}{\partial \rho} + b k^2 \right) \stackrel{!}{=} 0 \, .
\end{equation}
The transition from stability to instability occurs on the condition
\begin{equation}
 \rho_{\rm e} \left| \frac{\partial V_{\rm e}}{\partial \rho}
 \right| = b \rho_{\rm e} \tau k^2 + \left( 1  + \left| 
 \frac{\partial V_{\rm e}}{\partial V} \right| + \frac{\eta \tau k^2}{\rho} 
 \right) \left( \frac{a}{2} \pm \sqrt{ \frac{a^2}{4} + \frac{\partial {\cal
       P}_{\rm tot}}{\partial \rho}} \right) \, .
\end{equation}
In contrast to Eq.\ (\ref{instcond}), an additional instability appears
at large absolute wave numbers $|k| > 130$/km. 
However, these are connected with wavelengths that are
smaller than the average vehicle distance
$1/\rho$ so that they cannot be propagated by the 
discontinuous vehicular fluid. Perturbations which can actually be
propagated by the vehicles are moving in backward direction with
respect to $V_{\rm e}$ (cf.\ Fig.\ \ref{fig2}b).
\end{itemize}
\begin{figure}[htbp]
\unitlength1cm
\begin{center}
\begin{picture}(14,6.8)(0,-0.8)
\put(0,6){\epsfig{height=14cm, angle=-90, 
      bbllx=9cm, bblly=4.5cm, bburx=18.5cm, bbury=25cm, clip=,
      file=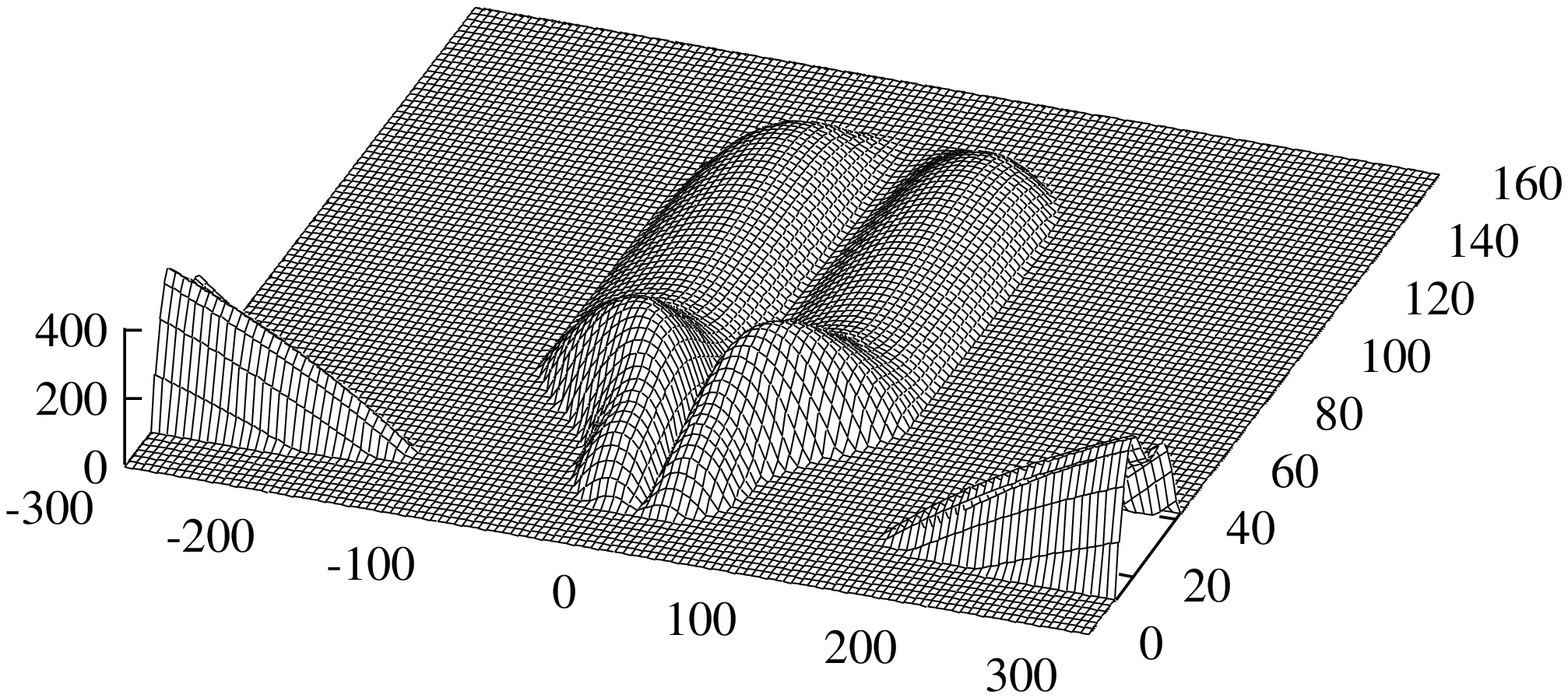}}
\put(1,5.8){\makebox(0,0){\large (a)}}
\put(1,4){\makebox(0,0){\large $\lambda(\rho_{\rm e},k)$ (1/h)}}
\put(7.5,-0.6){\makebox(0,0){\large $k$ (1/km)}}
\put(13,2.4){\makebox(0,0){\large $\rho_{\rm e}$}}
\put(13,1.8){\makebox(0,0){\large (veh/km)}}
\end{picture}
\begin{picture}(14,7.2)(0,-0.8)
\put(0,6.4){\epsfig{height=14cm, angle=-90, 
      bbllx=8.4cm, bblly=4.5cm, bburx=18.5cm, bbury=25cm, clip=,
      file=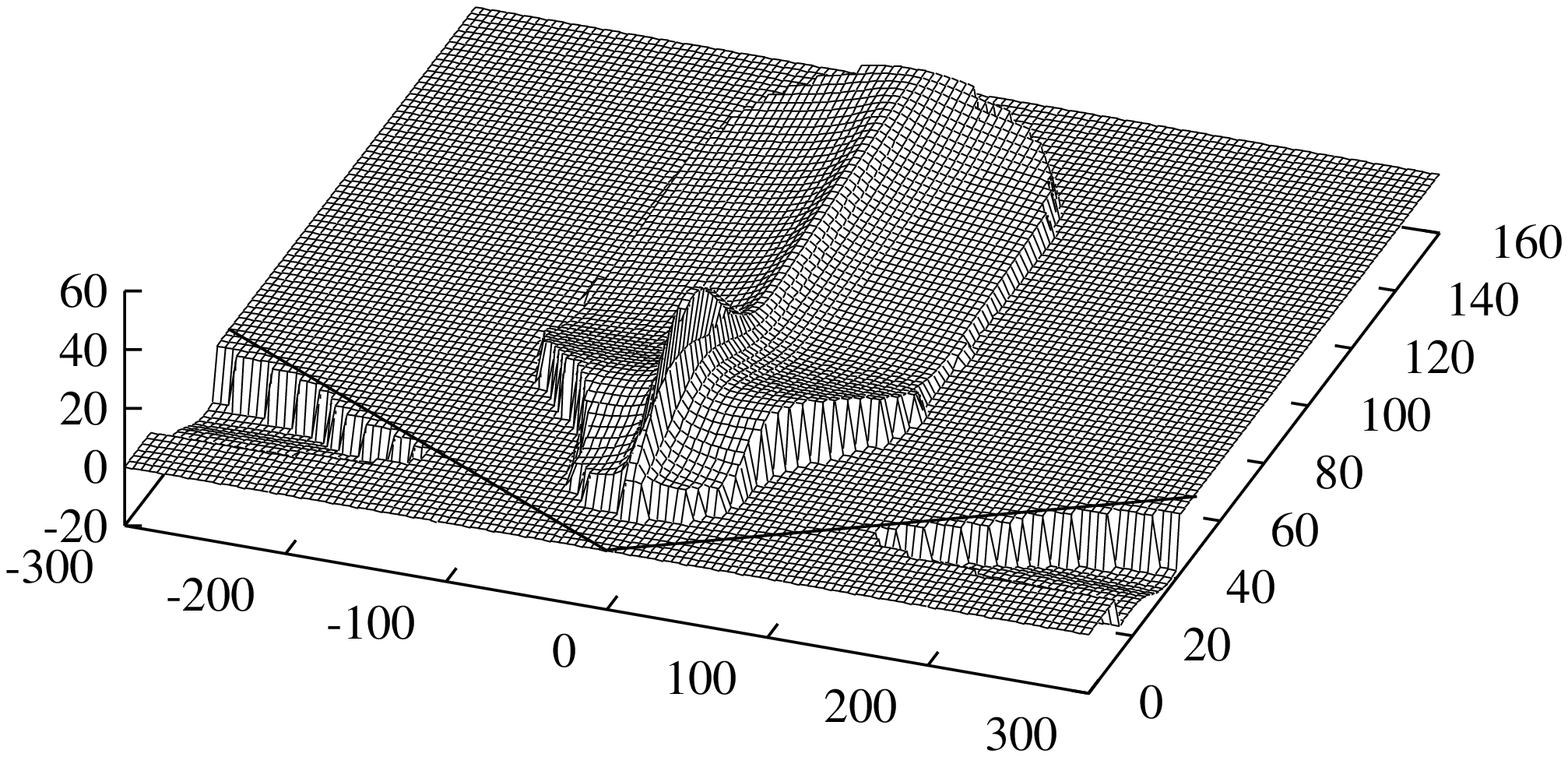}}
\put(1,6.2){\makebox(0,0){\large (b)}}
\put(1,5.2){\makebox(0,0){\large $-c(\rho_{\rm e},k)$}}
\put(1,4.5){\makebox(0,0){\large (km/h)}}
\put(7.5,-0.6){\makebox(0,0){\large $k$ (1/km)}}
\put(13,2.4){\makebox(0,0){\large $\rho_{\rm e}$}}
\put(13,1.8){\makebox(0,0){\large (veh/km)}}
\end{picture}
\end{center}
\caption{(a) Illustration of the largest growth parameter $\lambda$ 
for the instability region of traffic flow (i.e.\ where $\lambda(\rho_{\rm e},k) 
\ge 0$). In contrast to granular flow (cf.\ Fig.~\ref{fig1}), 
density waves develop at medium densities $\rho_{\rm e}$ and small absolute
wave numbers $|k| \ne 0$. %For $k=0$ we have {\em marginal} stability 
%($\lambda = 0$/h) due to the conservation of the number of vehicles. 
At moderate densities, the absolute wave number related to the largest growth 
rate increases with density, corresponding to a decrease of the
wave length of forming stop-and-go waves. As expected, traffic flow is only 
stable at low densities (free flow) and extreme densities (slow-moving 
traffic). The instabilities for large absolute wave numbers 
$|k| > 130$/km would be
connected with stop-and-go waves that move in forward direction with
group velocity $c(\rho_{\rm e},k) > 0$ 
relative to $V_{\rm e}(\rho_{\rm e})$ (b). However, these
instabilities are physically irrelevant, since they lie in front of the
lines $\pm 2 \pi \rho_{\rm e}$ which characterize the maximum wave numbers
that can be propagated by vehicles with an average distance of 
$1/\rho_{\rm e}$ (= minimum wave length).
%the discrete ``vehicular gas'' at density $\rho_{\rm e}$. 
The relevant instabilities at small absolute wave numbers
$|k| < 2 \pi \rho_{\rm e}$ are connected with stop-and-go waves that move
in backward direction with group velocity $|c(\rho_{\rm e},k)|
> 0$ relative to $V_{\rm e}(\rho_{\rm e})$. In contrast to Fig.
\protect\ref{kernkon}, the propagation speed has the right order of magnitude.}
\label{fig2}
\end{figure}

\section{Summary and Outlook}

It was shown that almost all macroscopic traffic models can
be viewed as special cases of the continuity equation and a certain
velocity equation. Although these equations have a close similarity to the 
hydrodynamic equations for ordinary fluids, they did not fulfill all 
consistency requirements for realistic traffic flow models. 
\par
Finally, it turned out
that the correct structure of fluid-dynamic traffic
equations looks considerably different. We obtained this result from
a refined version of Paveri-Fontana's gas-kinetic
traffic equation which was extended
by the effects of imperfect driving and vehicular space requirements.
In particular, we have applied the gas-kinetic theory of dense gases
and granular materials. 
\par
Nevertheless, despite the phenomenologically
similar behavior of traffic flow and granular material falling through
a narrow vertical pipe, the governing equations and instability mechanisms
are completely different. This was illustrated by numerical results of linear 
instability analyses and originates from the fact that momentum is conserved
by granular collisions but not by vehicular interactions.
\par
The parameters and relations occuring in the final model have been
estimated from empirical traffic data. In addition, we have empirically tested
the approximations that we made during the derivation of the fluid-dynamic
traffic model. Finally, it was shown that all consistency criteria are met.
\par
Future work will compare simulation results of the various discussed traffic
models. Moreover, the refined traffic flow model (which treats the 
highway lanes in an overall manner) will be extended to a model for the
different interacting highway lanes.

\subsection*{Acknowledgments}

The author is very grateful to H. Taale and the Dutch 
Ministry of Transport, Public
Works and Water Management for supplying the empirical traffic data.
He also wants to thank I. Goldhirsch, L. Schimansky-Geier, D. Rosenkranz,
M. Hilliges, V. Shvetsov, and R. K\"uhne for inspiring discussions.

%\end{multicols}
\end{document}